\documentclass[twocolumn]{aastex62}
\usepackage{xspace}
\usepackage[utf8]{inputenc}
\usepackage[T1]{fontenc}
\usepackage{ulem}
\usepackage{amsmath}

\definecolor{myblue}{rgb}{0.0, 0.5, 1.0}
\hypersetup{colorlinks=true,citecolor=myblue, linkcolor=myblue}

\shorttitle{Constraining HERA's in-situ beam pattern}
\shortauthors{Nunhokee et al.}

\begin{document}
	\title{Measuring HERA's primary beam in-situ: methodology and first results}
	
	\correspondingauthor{C.~D.~Nunhokee}
	\email{cnunhokee@berkeley.edu}
	
	\author{Chuneeta D. Nunhokee}
	\affiliation{Department of Astronomy, University of California, Berkeley, CA}
	
	\author{Aaron R. Parsons}
	\affiliation{Department of Astronomy, University of California, Berkeley, CA}
	
	\author{Nicholas S. Kern}
	\affiliation{Department of Astronomy, University of California, Berkeley, CA}
	
	\author{Bojan  Nikolic}
	\affiliation{Cavendish Astrophysics, University of Cambridge, Cambridge, UK}
	
	\author{Jonathan C. Pober}
	\affiliation{Department of Physics, Brown University, Providence, RI}
	
	\author{Gianni  Bernardi}
	\affiliation{INAF-Istituto di Radioastronomia, via Gobetti 101, 40129 Bologna, Italy}
	\affiliation{Department of Physics and Electronics, Rhodes University, PO Box 94, Grahamstown, 6140, South Africa}
	\affiliation{South African Radio Astronomy Observatory, Black River Park, 2 Fir Street, Observatory, Cape Town, 7925, South Africa}
	
	\author{Chris L. Carilli}
	\affiliation{National Radio Astronomy Observatory, Socorro, NM}
	
	\author{Zara  Abdurashidova}
	\affiliation{Department of Astronomy, University of California, Berkeley, CA}
	
	\author{James E. Aguirre}
	\affiliation{Department of Physics and Astronomy, University of Pennsylvania, Philadelphia, PA}
	
	\author{Paul  Alexander}
	\affiliation{Cavendish Astrophysics, University of Cambridge, Cambridge, UK}
	
	\author{Zaki S. Ali}
	\affiliation{Department of Astronomy, University of California, Berkeley, CA}
	
	\author{Yanga  Balfour}
	\affiliation{South African Radio Astronomy Observatory, Black River Park, 2 Fir Street, Observatory, Cape Town, 7925, South Africa}
	
	\author{Adam P. Beardsley}
	\affiliation{School of Earth and Space Exploration, Arizona State University, Tempe, AZ}
	
	\author{Tashalee S. Billings}
	\affiliation{Department of Physics and Astronomy, University of Pennsylvania, Philadelphia, PA}
	
	\author{Judd D. Bowman}
	\affiliation{School of Earth and Space Exploration, Arizona State University, Tempe, AZ}
	
	\author{Richard F. Bradley}
	\affiliation{National Radio Astronomy Observatory, Charlottesville, VA}
	
	\author{Jacob  Burba}
	\affiliation{Department of Physics, Brown University, Providence, RI}
	
	\author{Carina  Cheng}
	\affiliation{Department of Astronomy, University of California, Berkeley, CA}
	
	\author{David R. DeBoer}
	\affiliation{Department of Astronomy, University of California, Berkeley, CA}
	
	\author{Matt  Dexter}
	\affiliation{Department of Astronomy, University of California, Berkeley, CA}
	
	\author{Eloy  de~Lera~Acedo}
	\affiliation{Cavendish Astrophysics, University of Cambridge, Cambridge, UK}
	
	\author{Joshua S. Dillon}
	\affiliation{Department of Astronomy, University of California, Berkeley, CA}
	
	\author{Aaron  Ewall-Wice}
	\affiliation{Department of Physics, Massachusetts Institute of Technology, Cambridge, MA}
	
	\author{Nicolas  Fagnoni}
	\affiliation{Cavendish Astrophysics, University of Cambridge, Cambridge, UK}
	
	\author{Randall  Fritz}
	\affiliation{SKA-SA, Cape Town, South Africa}
	
	\author{Steve R. Furlanetto}
	\affiliation{Department of Physics and Astronomy, University of California, Los Angeles, CA}
	
	\author{Kingsley  Gale-Sides}
	\affiliation{Cavendish Astrophysics, University of Cambridge, Cambridge, UK}
	
	\author{Brian  Glendenning}
	\affiliation{National Radio Astronomy Observatory, Socorro, NM}
	
	\author{Deepthi  Gorthi}
	\affiliation{Department of Astronomy, University of California, Berkeley, CA}
	
	\author{Bradley  Greig}
	\affiliation{School of Physics, University of Melbourne, Parkville, VIC 3010, Australia}
	\affiliation{ARC Centre of Excellence for All-Sky Astrophysics in 3 Dimensions (ASTRO 3D)}
	
	\author{Jasper  Grobbelaar}
	\affiliation{SKA-SA, Cape Town, South Africa}
	
	\author{Ziyaad  Halday}
	\affiliation{SKA-SA, Cape Town, South Africa}
	
	\author{Bryna J. Hazelton}
	\affiliation{Department of Physics, University of Washington, Seattle, WA}
	\affiliation{eScience Institute, University of Washington, Seattle, WA}
	
	\author{Jacqueline N. Hewitt}
	\affiliation{Department of Physics, Massachusetts Institute of Technology, Cambridge, MA}
	
	
	\author{Daniel C. Jacobs}
	\affiliation{School of Earth and Space Exploration, Arizona State University, Tempe, AZ}
	
	\author{Austin  Julius}
	\affiliation{SKA-SA, Cape Town, South Africa}

	\author{Joshua  Kerrigan}
	\affiliation{Department of Physics, Brown University, Providence, RI}
	
	\author{Piyanat  Kittiwisit}
	\affiliation{School of Earth and Space Exploration, Arizona State University, Tempe, AZ}
	
	\author{Saul A. Kohn}
	\affiliation{Department of Physics and Astronomy, University of Pennsylvania, Philadelphia, PA}
	
	\author{Matthew  Kolopanis}
	\affiliation{School of Earth and Space Exploration, Arizona State University, Tempe, AZ}
	
	\author{Adam  Lanman}
	\affiliation{Department of Physics, Brown University, Providence, RI}
	
	\author{Paul  La~Plante}
	\affiliation{Department of Physics and Astronomy, University of Pennsylvania, Philadelphia, PA}
	
	\author{Telalo  Lekalake}
	\affiliation{SKA-SA, Cape Town, South Africa}
	
	\author{Adrian  Liu}
	\affiliation{Department of Astronomy, University of California, Berkeley, CA}
	\affiliation{Department of Physics and McGill Space Institute, McGill University, 3600 University Street, Montreal, QC H3A 2T8, Canada}
	\affiliation{Hubble Fellow}
	
	\author{David  MacMahon}
	\affiliation{Department of Astronomy, University of California, Berkeley, CA}
	
	\author{Lourence  Malan}
	\affiliation{SKA-SA, Cape Town, South Africa}
	
	\author{Cresshim  Malgas}
	\affiliation{SKA-SA, Cape Town, South Africa}
	
	\author{Matthys  Maree}
	\affiliation{SKA-SA, Cape Town, South Africa}
	
	\author{Zachary E. Martinot}
	\affiliation{Department of Physics and Astronomy, University of Pennsylvania, Philadelphia, PA}
	
	\author{Eunice  Matsetela}
	\affiliation{SKA-SA, Cape Town, South Africa}
	
	\author{Andrei  Mesinger}
	\affiliation{Scuola Normale Superiore, 56126 Pisa, PI, Italy}

	\author{Mathakane  Molewa}
	\affiliation{SKA-SA, Cape Town, South Africa}
	
	\author{Miguel F. Morales}
	\affiliation{Department of Physics, University of Washington, Seattle, WA}
	
	\author{Tshegofalang  Mosiane}
	\affiliation{SKA-SA, Cape Town, South Africa}
	
	\author{Abraham R. Neben}
	\affiliation{Department of Physics, Massachusetts Institute of Technology, Cambridge, MA}
	
	\author{Nipanjana  Patra}
	\affiliation{Department of Astronomy, University of California, Berkeley, CA}
	
	\author{Samantha  Pieterse}
	\affiliation{SKA-SA, Cape Town, South Africa}
	
	\author{Nima  Razavi-Ghods}
	\affiliation{Cavendish Astrophysics, University of Cambridge, Cambridge, UK}
	
	\author{Jon  Ringuette}
	\affiliation{Department of Physics, University of Washington, Seattle, WA}
	
	\author{James  Robnett}
	\affiliation{National Radio Astronomy Observatory, Socorro, NM}
	
	\author{Kathryn  Rosie}
	\affiliation{SKA-SA, Cape Town, South Africa}
	
	\author{Peter  Sims}
	\affiliation{Department of Physics, Brown University, Providence, RI}
	
	\author{Craig  Smith}
	\affiliation{SKA-SA, Cape Town, South Africa}
	
	\author{Angelo  Syce}
	\affiliation{SKA-SA, Cape Town, South Africa}
	
	\author{Nithyanandan  Thyagarajan}
	\altaffiliation{Jansky Fellow of the National Radio Astronomy Observatory}
	\affiliation{National Radio Astronomy Observatory, Socorro, NM} 
	\affiliation{School of Earth and Space Exploration, Arizona State University, Tempe, AZ}

	\author{Peter K.~G. Williams}
	\affiliation{Harvard-Smithsonian Center for Astrophysics, Cambridge, MA}
	
	\author{Haoxuan  Zheng}
	\affiliation{Department of Physics, Massachusetts Institute of Technology, Cambridge, MA}
 
\begin{abstract}
	The central challenge in 21~cm cosmology is isolating the cosmological signal from bright foregrounds. Many separation techniques rely on the accurate knowledge of the sky and the instrumental response, including the antenna primary beam. For drift-scan telescopes such as the Hydrogen Epoch of Reionization Array \citep[HERA, ][]{DeBoer2017} that do not move, primary beam characterization is particularly challenging because standard beam-calibration routines do not apply \citep{Cornwell2005} and current techniques require accurate source catalogs at the telescope resolution. We present an extension of the method from \citet{Pober2012} where they use beam symmetries to create a network of overlapping source tracks that break the degeneracy between source flux density and beam response and allow their simultaneous estimation. We fit the beam response of our instrument using early HERA observations and find that our results agree well with electromagnetic simulations down to a -20~dB level in power relative to peak gain for sources with high signal-to-noise ratio. In addition, we construct a source catalog with 90 sources down to a flux density of 1.4~Jy at 151~MHz.
\end{abstract}

\section{Introduction} 
\label{sec:intro}
The 21~cm line from neutral hydrogen (HI line) has gained attention as a probe of structure formation in the early Universe. Low-frequency observations probe high redshifts and therefore hold the potential to explore different epochs in the history of the Universe.
Using 21~cm HI line, we can study the Cosmic Dawn, when the first luminous sources such as Population III stars and massive X--ray sources formed  \citep{Furlanetto2016, McQuinn2016}, and the Epoch of Reionization (EoR), when these first luminous sources emitted ultraviolet and/or X--ray radiations and ionized the neutral intergalactic medium (IGM) over $z\sim$ 15--6 \citep{Ciardi2005, Zaroubi2013}. At closer redshifts, observations of the 21~cm line help probe the baryonic acoustic oscillations (BAO) which serve as tracers of the expansion of the Universe allowing study of the evolution of Dark Energy \citep{Bandura2014, Newburg2016}.

Considerable effort is being dedicated towards measuring highly redshifted 21~cm fluctuations. Active experiments include the Precision Array to Probe the Epoch of Reionization \citep[PAPER;][]{Parsons2010}, the Giant Metrewave Radio Telescope Epoch of Reionization \citep[GMRT;][]{Paciga2011}, the Murchison Widefield Array \citep[MWA;][]{Tingay2013}, the LOw Frequency Array \citep[LOFAR;][]{vanHaarlem2013}, the Canadian Hydrogen Intensity Mapping Experiment \citep[CHIME;][]{Bandura2014}, the Square Kilometer Array \citep[SKA;][]{Koopmans2015}, the Hydrogen Intensity and Real-time Analysis eXperiment \citep[HIRAX;][]{Newburg2016}, and HERA \citep{DeBoer2017}.

Observations of the 21~cm signal require the separation of astrophysical foregrounds that are $\sim$4--5 orders of magnitude brighter than the cosmological signal. Current foreground mitigation techniques take two flavors: foreground subtraction (or removal) and avoidance. Both techniques rely on the different spectral behavior of the foreground and the expected 21~cm emission. Subtraction methods attempt to model and subtract foreground emission from the observations \citep{Morales2006, Bowman2009, Liu2009, Bernardi2010, Bernardi2013, Chapman2016}, while the avoidance approach exploits how foregrounds are confined to a wedge--shaped region in the Fourier space to reveal an ``EoR window"  where the foreground contamination is minimal with respect to the 21~cm signal \citep{Morales2004, Vedantham2012, Liu2014a, Liu2014b, Pober2013, Thyagarajan2015a}. 

Recently, \citet{Kerrigan2018} showed that foreground modeling helps reduce foreground power in the wedge, mitigating its possible spillover in the EoR window. However, the efficiency of the subtraction method depends on accurate sky and instrument models: variable primary beams contribute to modeling errors, particularly for widefield observations. Though these errors can, in principle, be mitigated through calibration \citep{Mitchell2008, Bernardi2010, Bernardi2013, Yatawatta2013}, improper modeling can lead to additional errors that are significant enough to contaminate the cosmological signal \citep{Thyagarajan2015a, Thyagarajan2015b, Nunhokee2017, Procopio2017}.

Electromagnetic (EM) simulations are used to model primary beam responses, however, they fall short in including subtle real-world effects such as
feed misalignment, antenna-to-antenna variation and cross--coupling between neighboring antennas. Unfortunately, inaccurate beam models may lead to spurious spectral structures, improper scaling in the calibration solutions, and inaccurate power spectra normalization. Hence, we need techniques to characterize the instrument response using drift--scan widefield observations and validate outputs of the EM simulations.

Measurements of the primary beam is a requirement for both single dish radio telescopes and radio interferometers. Substantial efforts have been dedicated towards characterizing the primary beam of an antenna lately.
For example, \citet{Neben2016} and \citet{Line2018} use the ORBCOMM satellite to map the antenna response. 
Holography, a well-known technique in radio astronomy, is also used to study the instrument response \citep{Berger2016}.
Recently,  \citet{Virone2014, Pupillo2015, Jacobs2017, deLera2018} have demonstrated that we can make direct measurements of the primary beam using Unmanned Aerial Vehicles, however, they are accompanied by challenges that require further investigation.  Our work describes a technique to measure the instrument response using drift--scan widefield observations similar to \citet{Pober2012} and \citet{Eastwood2018}.

This paper is organized as follows: the formalism is derived in Section~\ref{sec:formalism}, the observations and data processing are described in Section~\ref{sec:observations}, and the results are discussed in Sections~\ref{sec:beam_measurements} and \ref{sec:source_cat}. 
We show the effects of confusion noise in Section~\ref{sec:confusion_noise} and conclude in Section \ref{sec:conclusions}.

\section{Formalism}
\label{sec:formalism}
\begin{figure}[ht!]
	\centering
	\includegraphics[width=0.5\textwidth]{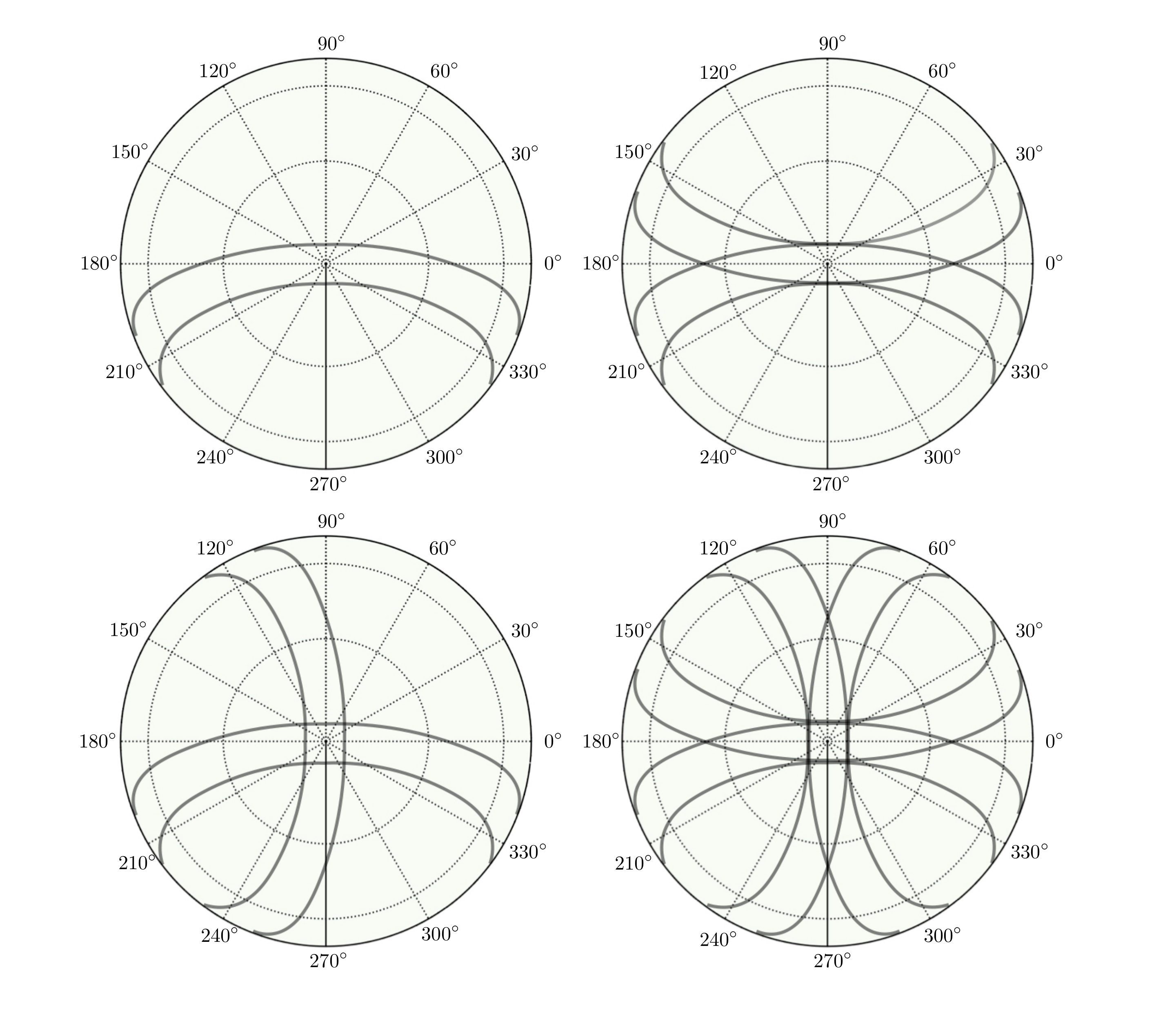}
	\caption{\textit{Top left}: A schematic representation of two arbitrary source tracks crossing the primary beam of our instrument. \textit {Top right}: Source tracks overlaid with the corresponding $180^{\circ}$ mirrored tracks. \textit{Bottom left}: Source tracks overlaid with the corresponding $90^{\circ}$ rotated tracks. \textit{Bottom right}: Source tracks overlaid with a combination of $180^{\circ}$ mirrored tracks, and $90^{\circ}$ and $270^{\circ}$ rotated tracks. The projection is orthographic with the center indicating zenith and the dotted lines are spaced $30^{\circ}$.}
	\label{fig:rotated_tracks}
\end{figure}

A zenith pointing telescope operating in drift--scan mode observes the radio sky as it passes across the primary beam of the instrument, placing astronomical sources at different points in the beam. The measured or apparent flux density of a source $I_{\nu}^{\prime}$ at any given time $t$ and frequency $\nu$ can be related to its intrinsic flux density $I_{\nu}$ as follows:
\begin{equation}\label{eq:corr_flux}
I_{\nu}^{\prime}(\hat{s}, t) = A_{\nu} (\hat{s}, t) I_{\nu}(\hat{s}) + n(\hat{s}, t)
\end{equation}
where $A_{\nu}$ is the primary beam response in the direction of the source $\hat{s}$ and time $t$ as sources pass through different beam responses at different time. Each measurement is associated with a noise denoted by $n(\hat{s}, t)$. 
Examples of two arbitrary sources being traced are shown in the top left panel of Figure~\ref{fig:rotated_tracks}.
If we know the intrinsic flux densities of these two sources, it is trivial to evaluate the primary beam values at these source locations using equation~\ref{eq:corr_flux}. Multiple sources then enable us to obtain full coverage of the instrument response.

However, precise measurements of the flux densities or catalogs consistent with the resolution of the observing telescope are often limited. We therefore simultaneously solve for $A_\nu(\hat{s}, t)$ and $I_{\nu}(\hat{s})$ in equation~\ref{eq:corr_flux}. We use a non-linear least square optimization as it has the ability to linearize the function and iteratively solves for the unknowns until convergence is achieved. Nevertheless, sources with different declinations transit through different parts of the primary beam, rendering the beam solutions degenerate with flux densities of the sources. We overcome this problem by introducing beam symmetries that break the degeneracy between the beam response and flux density. Following this approach, each source forms two or more tracks: the actual and the rotated/mirrored ones. Examples of beam symmetries are $180^{\circ}$ mirror symmetry (top right, Figure~\ref{fig:rotated_tracks}) and $90^{\circ}$ rotation (bottom left, Figure~\ref{fig:rotated_tracks}) about the source tracks. We can include multiple beam symmetries. For example, a combination of the $180^{\circ}$ mirror symmetry, and the $90^{\circ}$ and $270^{\circ}$ rotations about the source tracks, as shown in the bottom right panel of Figure~\ref{fig:rotated_tracks}. These overlapping source tracks relate to the flux densities of sources at different declinations.

\subsection{Forming least square problem}
\label{subsec:least_square}

We use a $n \times n $ grid 
to represent the empirical primary beam. Each grid pixel represents the sine projection of the azimuth--altitude angles $(\phi, \theta)$. The pixel resolution is estimated from the sampling such that each grid pixel within the FoV has at least one datum.
It must be small enough to preserve our assumption that the source tracks share the same beam response when they cross each other. To increase the Signal--to--Noise Ratio (SNR), we represent the primary beam value at any given source location and time using a linear interpolation of the four closest pixels.

Now that we have our gridded measurements and a network of overlapping source tracks using the beam symmetries, we form a least square optimization problem. For any given source $\hat{s}$, we aim to minimize
\begin{align}\label{eq:chisquare}
&{\rm min } \sum_{t, \hat{s}}\Big|I_{\nu}^{\prime} (\hat{s}, t) - A_{\nu} (\hat{s}, t)I_{\nu} (\hat{s})\Big|^2\nonumber\\
=&{\rm min } \sum_{t, \hat{s}}\Big|I_{\nu}^{\prime} (\hat{s}, t) - \sum \limits_{k=0}^4 a_{k}(\hat{s}, t)\frac{1}{d_k}I_{\nu} (\hat{s})\Big|^2
\end{align}
where the summation over $k$ represents the linear interpolation between the four closest pixels with $a_k$ denoting the beam solution at grid index $k$. Each grid index $k$ is mapped to a $(\phi, \theta)$ on the 2--dimensional grid. The distance $d_k$ is calculated with respect to the azimuth--altitude angle $(\phi_0, \theta_0)$ evaluated at source $\hat{s}$ and time $t$.

With multiple sources, we can construct a solvable matrix in the form
\begin{equation}\label{eq:nlinear_sys}
{\bf m} = {\bf C}{\bf \hat{x}} + {\bf \hat{n}}
\end{equation}
where ${\bf m}$ is the vector of measurements containing values for $I_{\nu}^{\prime}(\hat{s}, t)$, ${\bf C}$ is the condition matrix describing the combination and conditions of the parameters, and ${\bf \hat{x}}$ contains the unknown parameters: beam response $a_k$ and source flux density $I_{\nu}(\hat{s})$. 

Given that the system shown in equation~\ref{eq:nlinear_sys} comprises two unknowns parameters, it is non-linear, hence we use the Gauss-Newton algorithm to solve for $a_k$ and $I_{\nu}(\hat{s})$. This algorithm is second--order iterative whose convergence is dependent on the initial guesses of the unknown parameters. At each iteration, the equation is expressed as a Taylor expansion about the solutions derived in the previous iteration using the \verb linsolve \footnote{\url{https://github.com/HERA-Team/linsolve}} package.

\subsection{Using prior knowledge of the sky}
\label{sec:sky_prior}
As mentioned in Section~\ref{subsec:least_square}, we need to provide initial guesses for a better convergence in equation~\ref{eq:chisquare}. Estimates of $I_{\nu}(\hat{s})$ are calculated using
\begin{equation}\label{eq:corr_fflux}
I_\nu(\hat{s}) = \frac{\sum\limits_t w_\nu(\hat{s},t) I_\nu^{\prime} (\hat{s}, t)}{\sum\limits_t w_\nu(\hat{s},t)A^{{\rm M}}_{\nu}(\hat{s},t)}
\end{equation}
where $A^{{\rm M}}_{\nu}(\hat{s},t)$ is the primary beam value evaluated from EM simulations at source $\hat{s}$, time $t$ and frequency $\nu$, and $w_\nu$ are the weights assigned to the measurements $I_\nu^{\prime}$. In our case, we take $w_\nu$ to be $A^{{\rm M}}_{\nu}(\hat{s},t)$. 
Given that $I_\nu^{\prime}$ are already weighted by the instrument's primary beam, this additional weighting accounts for the time--samples averaged into each beam pixel, thus, providing an effective approximation of the inverse-variance weighting \citep{Jacobs2013}.

\subsection{Electromagnetic simulations}
\label{sec:electromagnetic_sims}
\begin{figure}[!ht]
	\centering
	\includegraphics[width=0.45\textwidth]{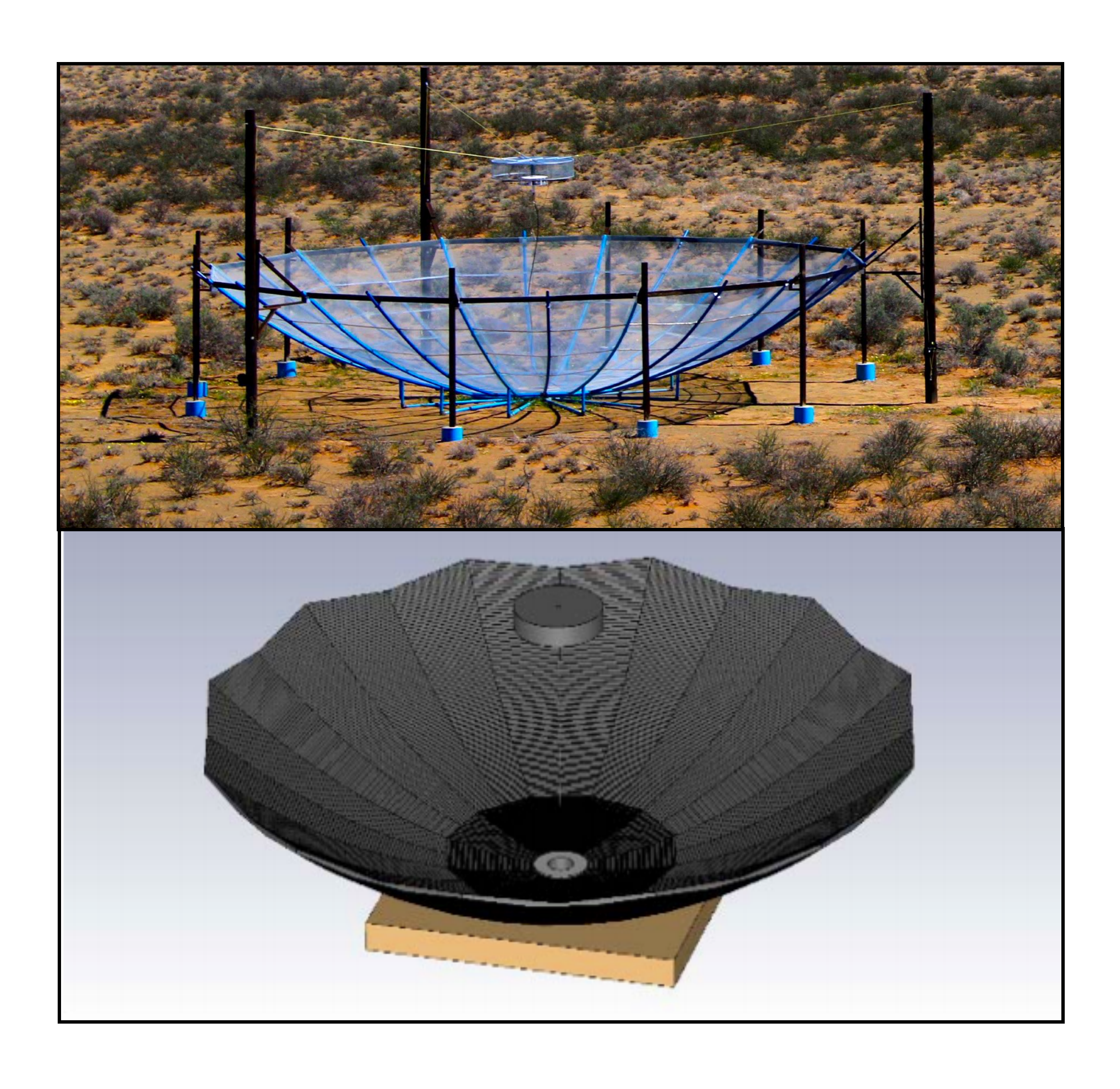}
	\caption{{\textit Top}: A HERA dish with a crossed-dipole feed as installed at Karoo desert in South Africa. \textit{Bottom}: A 14 m dish and a crossed-dipole feed model generated using CST for EM simulations.}
	\label{fig:hera_dish_setup}
\end{figure}

The beam models used in this work are generated using the Computer Simulation Technology (CST) software \citep{Fagnoni2019}. The antenna setup used for the simulation is illustrated in Figure~\ref{fig:hera_dish_setup}. 
The dish is a 14~m diameter paraboloid made up of aluminum mesh, held by a framework of PVC pipes converging into a cylindrical concrete slab at the center of the dish. The crossed-dipole feed (previously used for PAPER) is enclosed in a cylindrical cage.
It is sensitive to East--West and North--South polarizations, and is sandwiched between two steel discs that act as sleeves to broaden the frequency response by creating a dual-resonance structure \citep{Parsons2010}. 

\begin{figure}[ht!]
	\centering
	\includegraphics[width=0.5\textwidth]{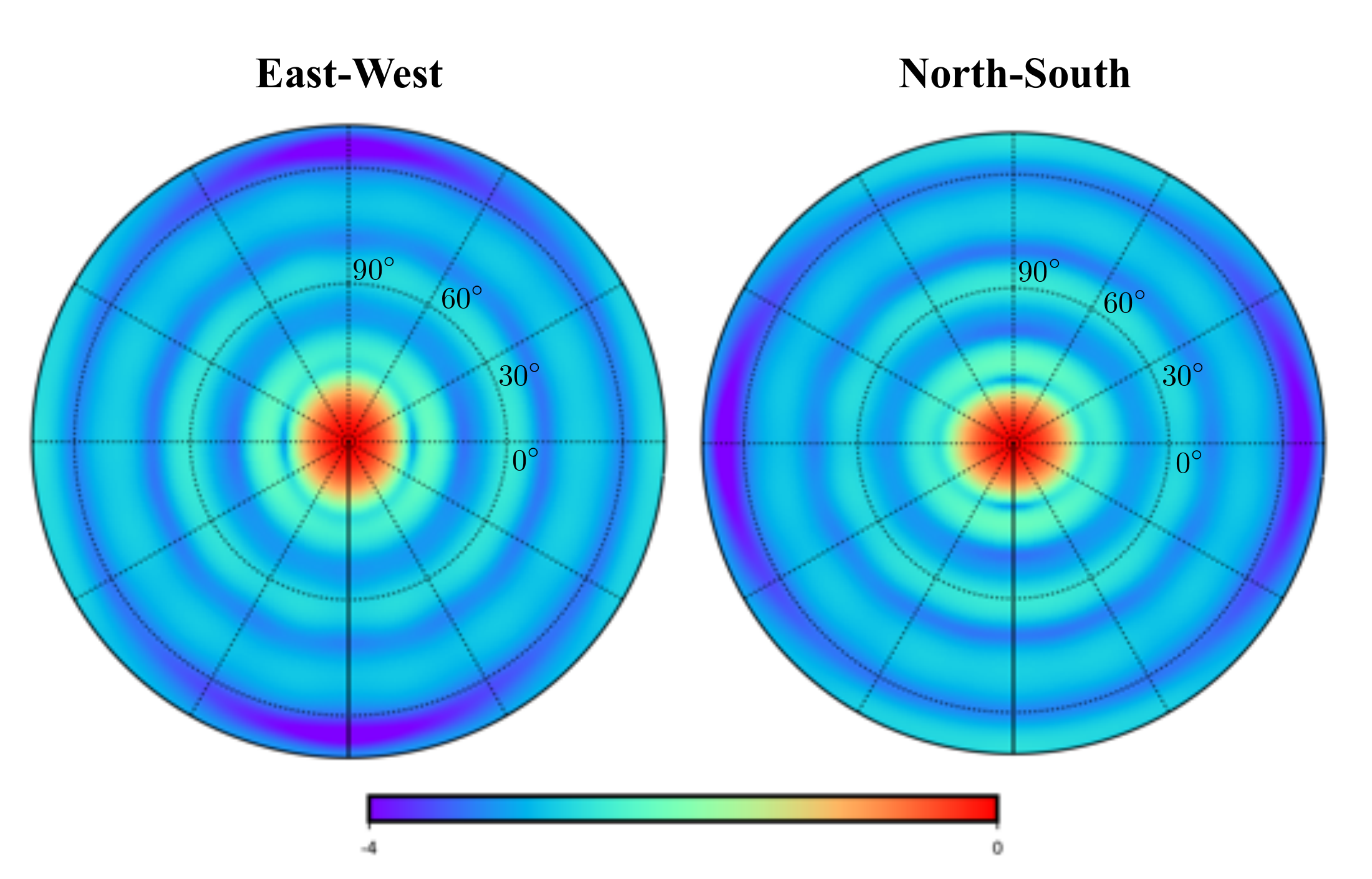}
	\caption{Primary beam models generated by the EM simulation in the East--West (left) and North--South (right) polarizations.}
	\label{fig:simulated_beam}
\end{figure}

The simulation accounts for the EM properties of different materials in use, including conductivity of metals, dielectrics of the propagation medium, and signal attenuation \citep{Fagnoni2019}. However, they do not include certain real-world effects such as uneven ground plane, and feed misalignment. \citet{Fagnoni2019} produce beam models both with and without the effects of mutual coupling. We use the EM simulations without mutual coupling. The resulting primary beam models for the East--West and North--South polarizations can be seen in Figure~\ref{fig:simulated_beam}. The beam model in the East--West polarization is assumed to be a $90^{\circ}$ rotation of the North--South polarization.

\section{Observation}
\label{sec:observations}

\begin{figure}[!ht]
	\hspace{-0.6cm}
	\includegraphics[width=0.5\textwidth]{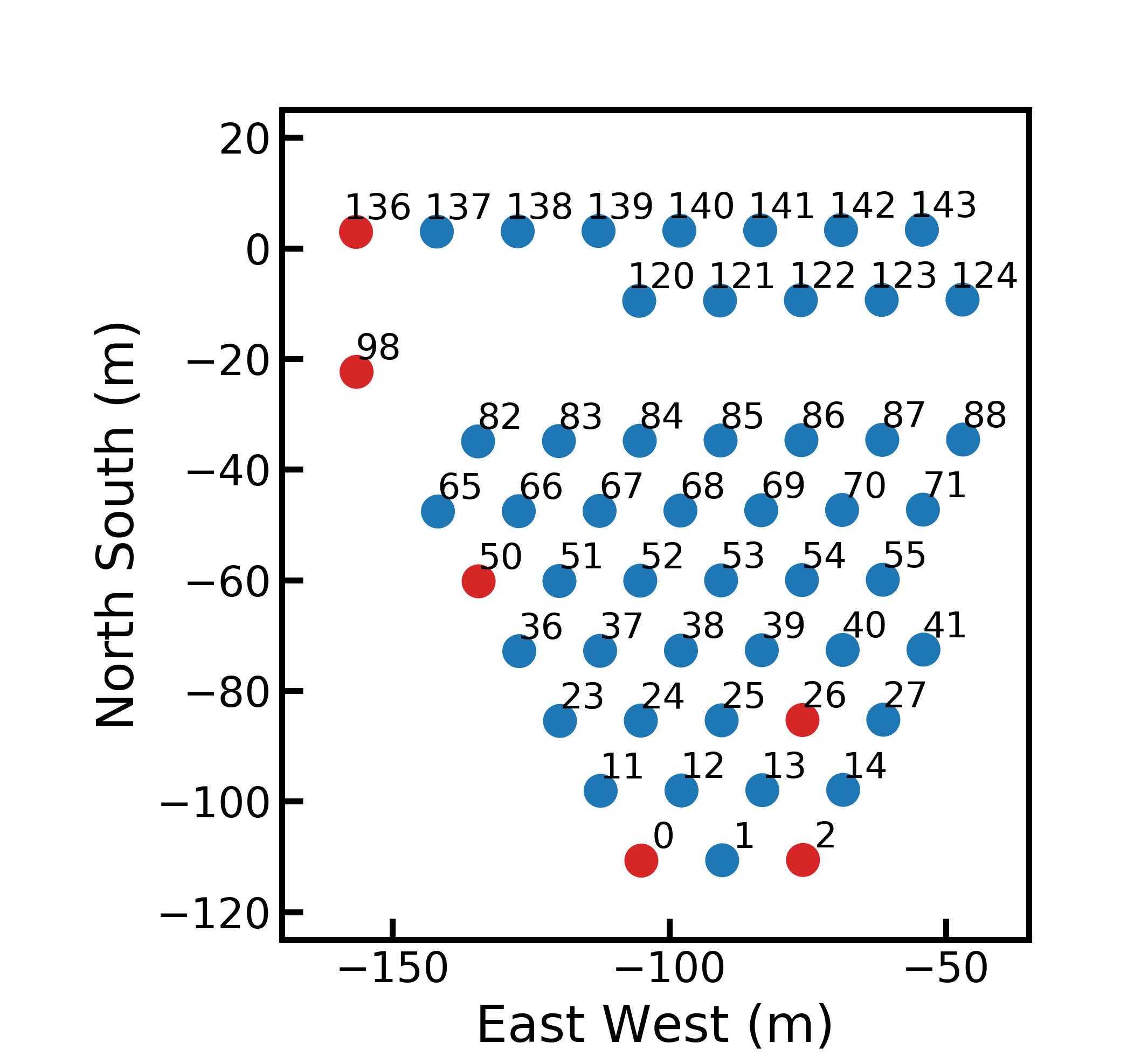}
	\caption{HERA-52 antenna configuration. Malfunctioning elements highlighted in red for Dec 24, 25, and 26, 2018 were not included in this work.}
	\label{fig:hera_layout}
\end{figure}

\begin{figure}
	\centering
	\includegraphics[width=0.5\textwidth]{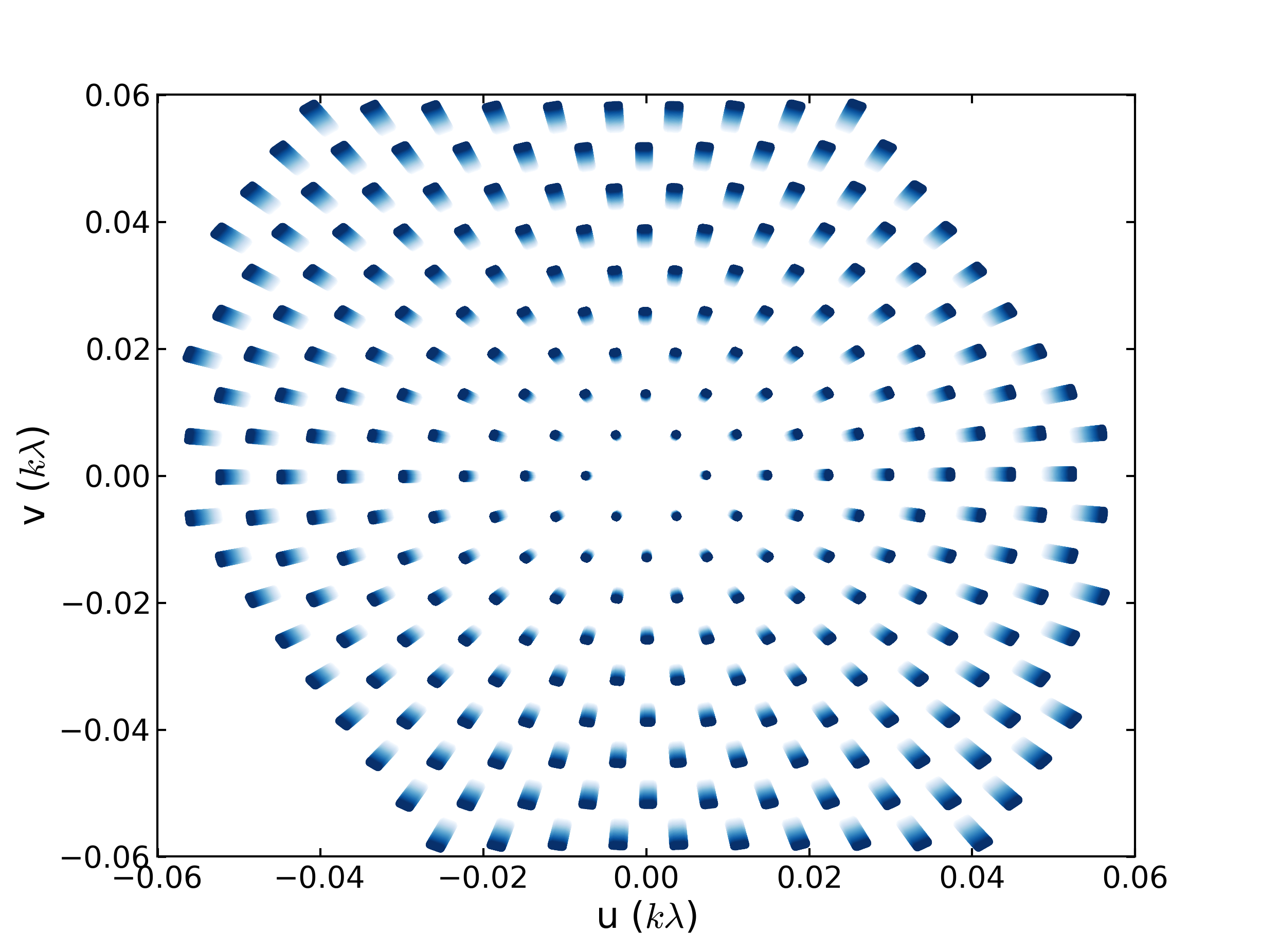}
	\caption{UV sampling generated from the antenna configuration for a 10~MHz frequency band centered at 150~MHz. The transition in color from faint to bright represents the variation of the uv sampling as a function of increasing frequency. Note the change in amplitude as sources cross through the primary beam.}
	\label{fig:uv_coverage}
\end{figure}

The measurements used in this paper are taken from HERA Phase I configuration, comprising 52 antennas (Figure~\ref{fig:hera_layout}) that reach out to a $uv$ distance of 75~$\lambda$ as illustrated in Figure~\ref{fig:uv_coverage}. Details of the observational setup are listed in Table~\ref{tab:observation_setup}. 

\begin{table}[ht]
	\caption{Summary of HERA Phase I setup}
	\centering
	\begin{tabular}{m{12em}m{10em}}
		\hline \hline
		Array longitude & 21$^{\circ}$25$^{\prime}$41.9$^{\prime \prime}$ \\
		Array latitude & -30$^{\circ}$43$^{\prime}$17.5$^{\prime \prime}$\\
		Number of dishes & 52\\
		Frequency range &  100--200~MHz\\
		Number of channels & 1024\\
		Integration time & 10.7~s\\
		Daily Observing time & 6~pm to 6~am SAST\\
		Daily Observing duration & 12~hours\\
		Observing days & Dec 24, 25, 26, 2018 \\[1ex]
		\hline
	\end{tabular}
	\label{tab:observation_setup}
\end{table}

\begin{figure*}[h!tp]
	\centering
	\includegraphics[width=0.98\linewidth]{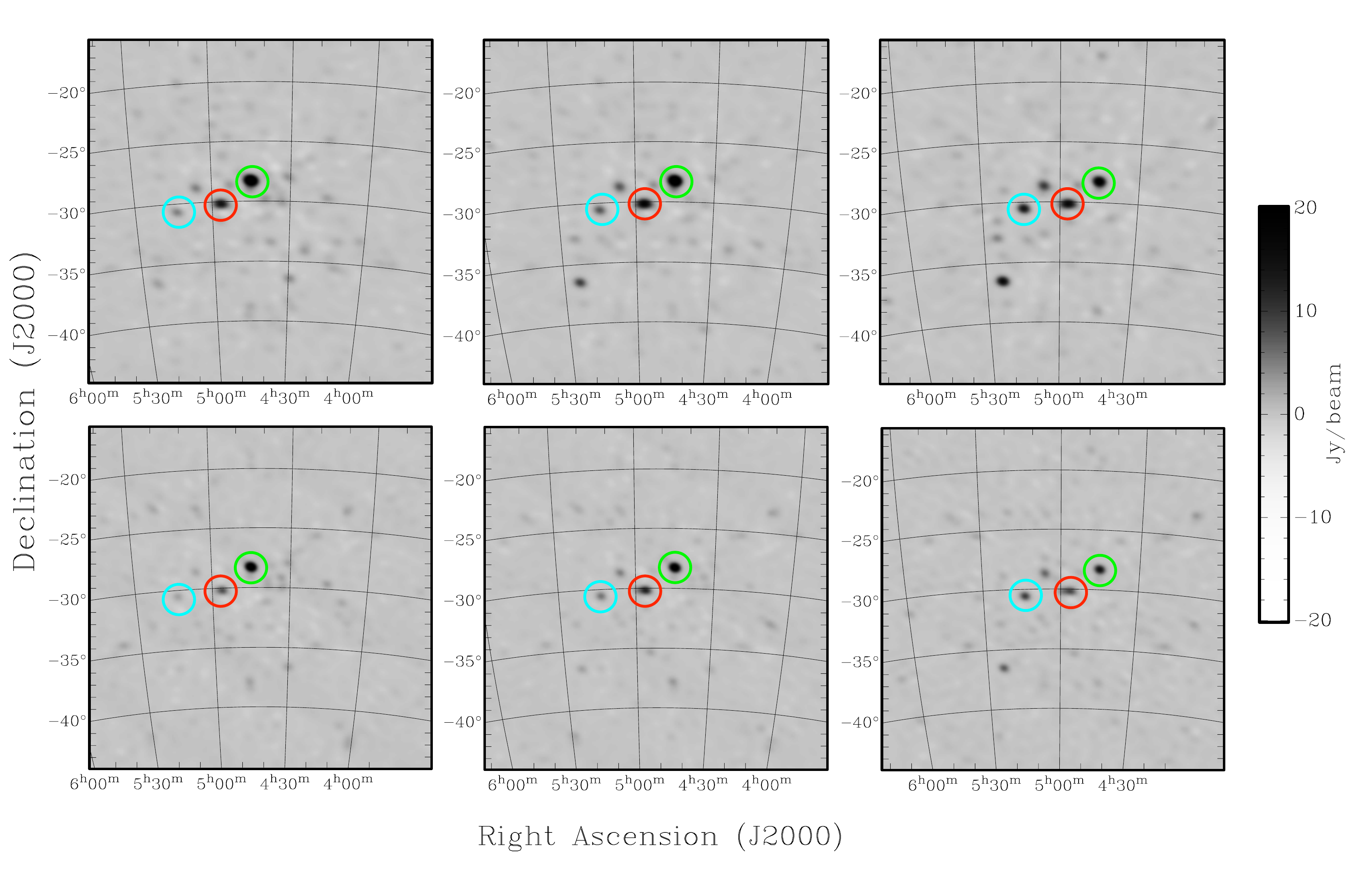}
	\caption{Snapshot images formed from 10~MHz band centered at 130~MHz (top) and 150~MHz (bottom) for three consecutive 10--minute observations for the East--West polarizations.  The three brightest sources in the snapshots are highlighted in red, blue and green.}
	\label{fig:msf_images}
\end{figure*}

\subsection{Data Reduction and calibration}
\label{sec:data_processing}
We use data from the Second Internal Data Release (IDR2) processed with the HERA analysis and reduction pipeline, which are flagged for faulty antennas and Radio Frequency Interference (RFI) \citep{Kerrigan2019, Beardsley2019}.

The HERA layout is such that we can employ redundant calibration, a powerful technique that uses the property of redundant baselines measuring the same sky-signal \citep{Zheng2014, Liu2010, Dillon2016, Dillon2020}.
Redundant calibration iteratively solves for the per antenna, per time and per frequency instrumental complex gains by minimizing the sum of the deviations of the per baseline  measured visibilities $V_{ij}^{\prime}$ from that of the model $V_{ij}(t,\nu)$ such that
\begin{equation}\label{eqn:redcal}
\rm{min}\sum_{bl}\frac{|V_{ij}^{\prime} - g_ig_j^*V_{ij}(t,\nu)|^2}{\sigma_{ij}^2}
\end{equation}
where $g_i$ and $g_j$ are the complex antenna gains for antenna $i$ and $j$ respectively, and $\sigma^2_{ij}$ is the noise in the per baseline measured visibilities. The sum is taken over all the redundant baselines denoted by $bl$. The model visibilities $V_{ij}$ are initially taken as the averaged visibilities of the redundant baselines and are updated at each iteration.

However, the overall amplitude and phase of the complex gain solutions obtained from equation~\ref{eqn:redcal} remain unknown. In order to correct for these overall amplitude, and phase degeneracies the data must be tied to a model of the sky \citep{Dillon2018}. In the HERA Phase I reduction pipeline this is done using bright sources from the GLEAM catalog \citep{Hurley-Walker2017} in conjunction with the EM simulations without mutual coupling \citep{Fagnoni2019} to create a set of model visibilities that is used to solve for direction-independent antenna-based gains for a single field \citep{Kern2020}, achieved using standard routines in the Common Astronomy Software Applications (CASA) package. This calibration is applied to one night of good data, which themselves then become the model visibilities for calibrating other nights in the data release. For more details on redundant and absolute calibration in the HERA reduction pipeline we refer the reader to \citet{Dillon2020} and \citet{Kern2020}, respectively. 

\subsection{Imaging}
\label{sec:imaging}
Radio interferometers propagate voltages from each antenna to a correlator where the inputs are cross-multiplied and time--averaged. These cross-correlated measurements, termed as visibilities are Fourier transforms of the radio emission from the sky under a flat--sky approximation. Hence, to image the sky, we need to Fourier transform the data back to the image domain. The operation is not straightforward as the measured visibilities are now convolved with the instrument response. Assuming a flat--sky, the resulting sky image $I^D(l, m, \nu)$ can be obtained using \citep{Thompson2017}

\small
\begin{equation}\label{eqn:dirty_image}
A(l, m, \nu)I^D(l, m, \nu) = \int\int S(u,v)V^c(u, v, \nu) e^{2\pi i (ul + vm)}~dudv
\end{equation}
\normalsize
where $V^c(u,v)$ denotes the calibrated visibilities (refer to Section~\ref{sec:data_processing}),  $S(u,v)$ represents the sampling function (also known as the $uv$ coverage; Figure~\ref{fig:uv_coverage})) and $(l, m)$ are the direction cosines relative to the source position. The resulting image $I^D$ is generally known as ``the dirty image" as it is the sky signal convolved with the point spread function (Fourier transform of the sampling function).

Widefield imaging is limited by various factors including bandwidth and time smearing, and non--coplanar baselines \citep{Cornwell2005}. While forming images, we need to choose the time interval and bandwidth over which we can assume constant source emission to prevent radial smearing. Single-channel imaging is ideal, however, given the poor $uv$-coverage of our instrument we need to find an optimal frequency interval with minimum bandwidth smearing. This optimal frequency band $\Delta \nu$ is dependent on the resolution of the instrument such that \citep{McMullin2007}
\begin{equation}\label{eq:band_smearing}
\Delta \nu < \nu_0 \bigg(\frac{D}{b_{\text{max}}}\bigg)
\end{equation}
where $b_{\text{max}}$ is the maximum baseline length, $D$ is the dish diameter and $\nu_0$ is the reference frequency (taken as the center frequency in our case). We find that assuming constant source emission over a 10~MHz band and 10~minutes interval provides us with minimum smearing along time and frequency axes.

To form images, we use a sampling of $\Delta l \approx \frac{1}{2u_{\text{max}}} \approx 8^{\prime}$ and  $\Delta m \approx  \frac{1}{2v_{\text{max}}} \approx 8^{\prime}$ and the multi-frequency synthesis (MFS) algorithm embedded in the CASA package. The MFS algorithm combines data from all the frequency channels within the specified band onto a single spatial--frequency grid, assuming a constant sky-brightness throughout. This assumption can cause spurious spatial structures for sources with spectral variations and high dynamic ranges. Since our current observations are looking at a dynamic range of a few hundreds, the basic MFS algorithm is sufficient. 

Another limitation of widefield imaging is that the standard assumption of non-zero $w$--term in interferometric imaging (equation~\ref{eqn:dirty_image}) no longer applies and, hence sources that are away from zenith may be distorted or may introduce artefacts \citep{Cornwell2005}. We use the $w$--projection algorithm in CASA to correct for the $w$--terms. 
We choose 128 $w$--projections based on HERA's FoV which is about 20$^{\circ}$. 

Furthermore, the images are formed using baselines greater than 30~m to reduce contaminations from diffuse emission. The data are then uniformly gridded onto the $uv$--plane, boosting image pixels with low weights, thus allowing for a sharp resolution and sidelobe reduction in the FoV. Each grid pixel is weighted by the inverse of the number of visibilities that fall within the grid. As the generated images are convolved with the point spread function (also known as the ``dirty beam"), we use the iterative Cotton-Schwab deconvolution algorithm \citep{Cornwell2005}, embedded in CASA to isolate the sky image from the point spread function. The cleaning mask is set to the full width half maximum (FWHM) of the primary beam and clean down to the first negative clean component. Deconvolved images centered at 130~MHz (top panel) and 150~MHz (bottom panel) are shown in Figure~\ref{fig:msf_images}. They comprise mostly point sources and the dynamic range is $~\sim 1:30$. The size of these point sources decreases as a function of increasing frequency as a result of variation in the point spread function. The flux density as well decreases with increasing frequency as extragalactic sources emit synchrotron emission.

\subsection{Source Extraction}
\label{sec:source_extraction}

\begin{figure}[h!]
	\centering	
	\includegraphics[width=0.95\linewidth]{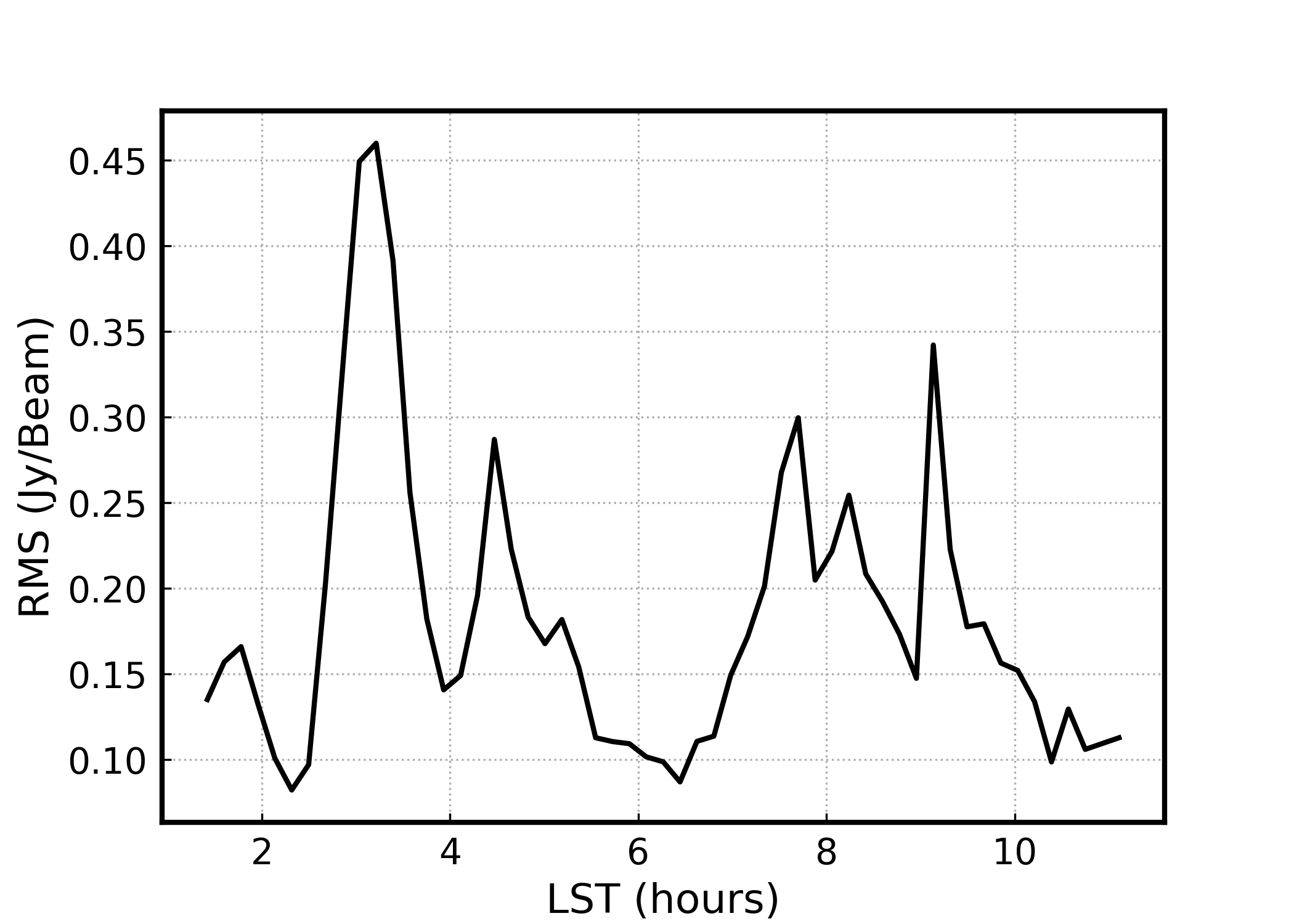}
	\caption{RMS of the confusion background obtained by fitting a Gaussian profile to the flux density distribution in the images generated at different LSTs. The average RMS is about 0.18~Jy, thus the threshold for source selection is chosen to be about five times the RMS value.}
	\label{fig:image_rms}
\end{figure}

\begin{figure}[h!]
	\centering
	\includegraphics[width=1.1\linewidth]{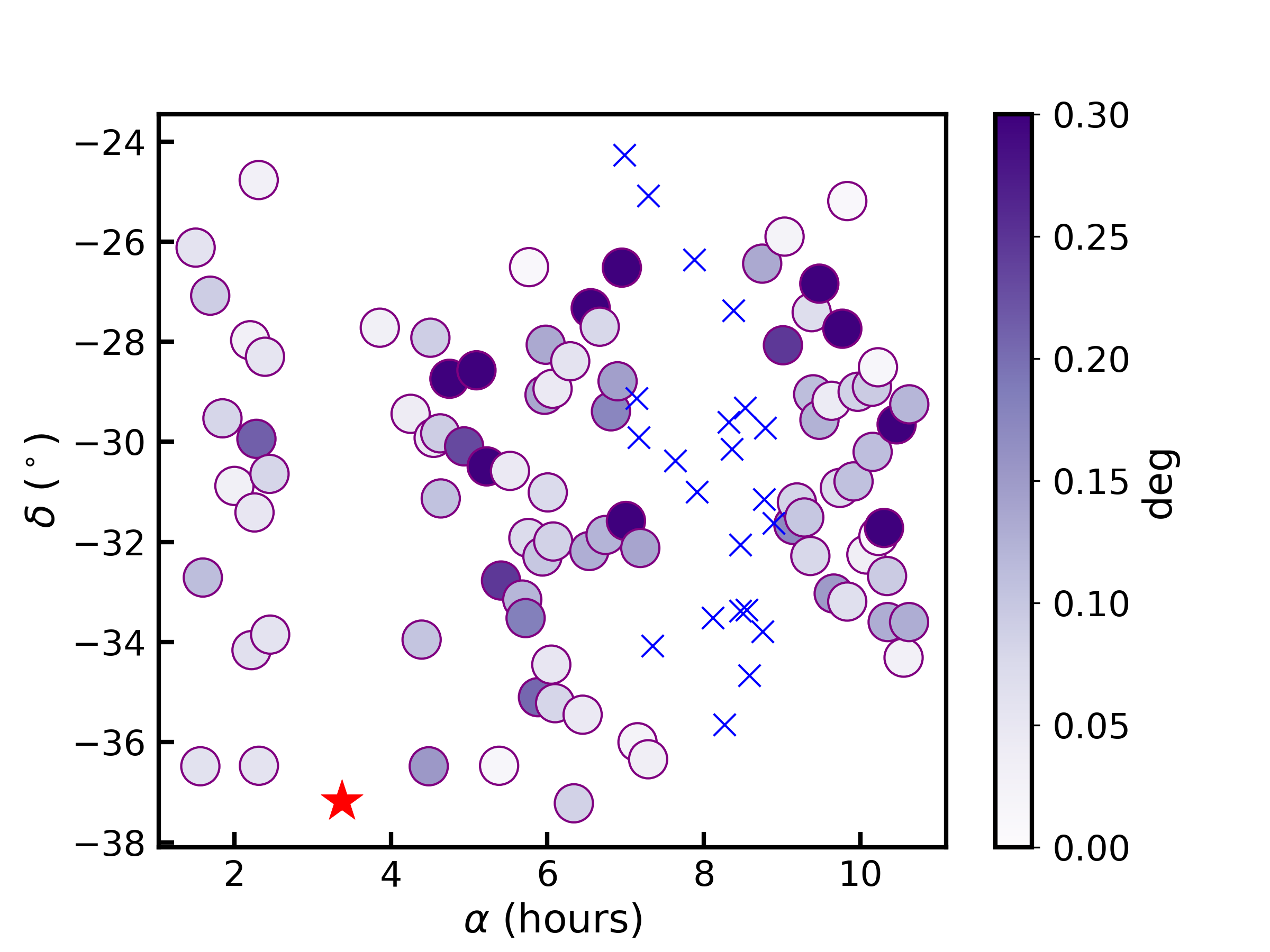}
	\caption{HERA sources found in the GLEAM survey are plotted in circles. Color indicates the absolute difference in angular position between the sources identified by PyBDSF and the matched candidates from the GLEAM survey. The blue crosses denote the sources that are not included in the GLEAM survey. The red star represents the brightest source in the FoV, Fornax~A.}
	\label{fig:hera_gleam_pos}
\end{figure}

\begin{figure*}[h!tp]
	\centering	
	\includegraphics[width=0.95\linewidth]{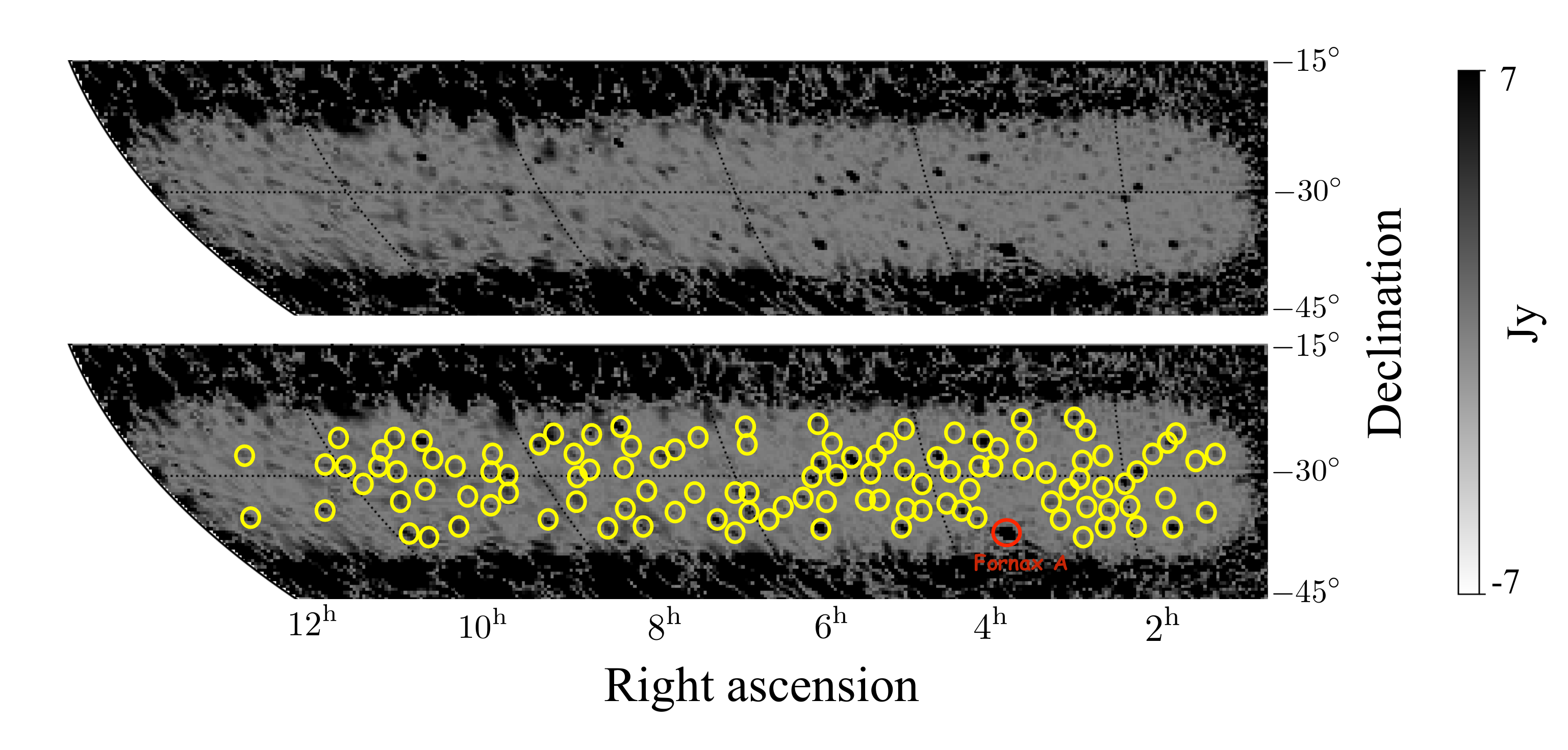}
	\caption{\textit{Top}: A mosaic map formed from 10~minute observations spanning 12~hours in LST. \textit{Bottom}: The same map with yellow circles marking the 112 potential source candidates identified by PyBDSF. The brightest source, Fornax~A is highlighted in red.}
	\label{fig:healmap}
\end{figure*}

Once we have the deconvolved images as described in Section~\ref{sec:imaging}, we run source finder software Python Blob Detection and Source Finder \citep[PyBDSF,][]{Mohan2015} to find potential point sources. The finder identifies source candidates in islands of pixels brighter than 1~Jy. This choice of threshold is estimated using the root mean square (RMS) across pixels in the images generated at different local sidereal times (LST). We fit a Gaussian to the pixel intensity distribution and calculate the standard deviation.  Figure~\ref{fig:image_rms} shows the confusion noise as a function of the LST. We notice that the confusion noise peaks when a bright source is in or near the FoV, for instance, we have Fornax~A, Pictor~A, Virgo~A and Centaurus~A transiting at LST 3, 5, 12 and 13~hours respectively. The average RMS across LSTs is about $\sim$~0.18~Jy, hence we set PyBDSF threshold to be five times the RMS value, identifying sources above 1~Jy.

Since each image spans a 10--minute observation and the beam--crossing timescale is 10~minutes, the sources can be found in different images, possibly with different centered peaks. We overcome this issue by averaging all sources within 1~arcmin resulting in 113 unique sources including Fornax~A. The declination range of these sources is limited within HERA's FoV between $-20^{\circ}$ to $-40^{\circ}$ while the right ascension spans the duration of the observation which is 12~hours.

We then compare our extracted source candidates with the GLEAM survey \citep{Hurley-Walker2017}. 
We find counterparts for 90 of the sources, plotted in Figure~\ref{fig:hera_gleam_pos}. 
Most of the extracted positions match with the GLEAM results within 10--20$\%$ except a few sources for which the difference goes up to 30$\%$ (refer to Section~\ref{sec:source_cat} for a detailed explanation). 
The remaining 23 sources include Fornax~A (red star; Figure~\ref{fig:hera_gleam_pos})) and sources located within the Magenallic Clouds (blue crosses; Figure~\ref{fig:hera_gleam_pos})) that are excluded in the GLEAM survey.

We produce a mosaic (Figure~\ref{fig:healmap}) using the individual snapshots to illustrate the sky seen by HERA at 150~MHz during a night of observation. The mosaic is constructed using equation~\ref{eq:corr_fflux}, hence it shows estimates of the intrinsic flux densities $I_{\nu}(\hat{s})$. We project the deconvolved images onto a HEALPIX grid \citep{Gorski2005} and the resulting maps are inputs to $I_\nu^{\prime}(\hat{s}, t)$, and $A_\nu^M(\hat{s}, t)$ is calculated from the EM simulations. 
The top panel of Figure~\ref{fig:healmap} depicts the 20$^{\circ}$ FoV of HERA. Our measurements are dominated by noise at the edges and beyond the FoV resulting in noisy $I_{\nu}(\hat{s})$.
We highlight the source candidates identified by PYBDSF in yellow in the bottom panel of Figure~\ref{fig:healmap}. The brightest source Fornax~A is circled in red.

\begin{figure*}
	\centering
	\includegraphics[width=0.8\textwidth]{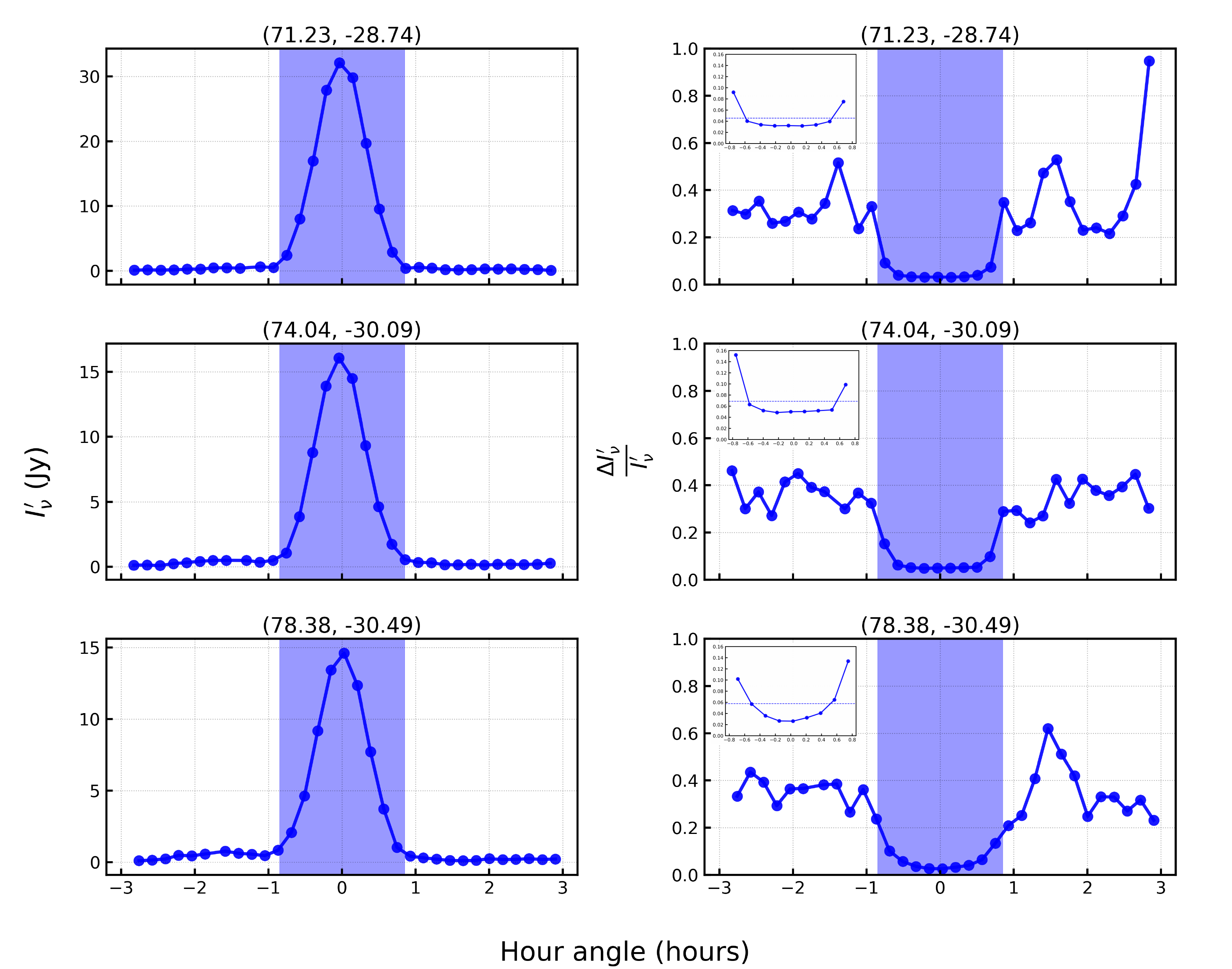}
	\caption{\textit{Left panel}: Flux densities of the sources marked in blue (top), red (middle) and green (bottom) in Figure~\ref{fig:msf_images} measured at 150~MHz. The shaded region indicates the time span the source transits the main lobe of the instrument's primary beam. \textit{Right panel}: Relative error associated with the measured flux densities. The inlet plot at the top left corner shows a zoom version of the relative error for the source crossing the main lobe and the dotted line represents the corresponding average relative error.}
	\label{fig:src_tracks}
\end{figure*}

Given the locations (right ascensions and declinations) of the source candidates, we extract their flux densities as follows. For a given source position $(\alpha, \delta)$:
\begin{enumerate}
	\item we select a region centered at the source location, with a radius equal to the synthesized beam.
	\item we fit a Gaussian distribution to the selected region and evaluate the integrated flux density $I_{\nu}^{\prime}$. We prefer the results of the Gaussian fit over PyBDSF outputs because of the gaps in our $uv$--coverage resulting in a poor performance of PyBDSF.
	\item we subtract the Gaussian distribution from the data and select a region, centered at the source location, with a radius equal to twice the synthesized beam. We then calculate the standard deviation of the selected region  to obtain an estimate of the error $\Delta I_{\nu}^{\prime}$ involved in the extraction process. The reason for choosing twice the synthesized beam for the error calculation is to include sidelobe contribution from nearby sources. 
\end{enumerate}

\begin{figure}[ht!]
	\centering	
	\includegraphics[width=0.95\linewidth]{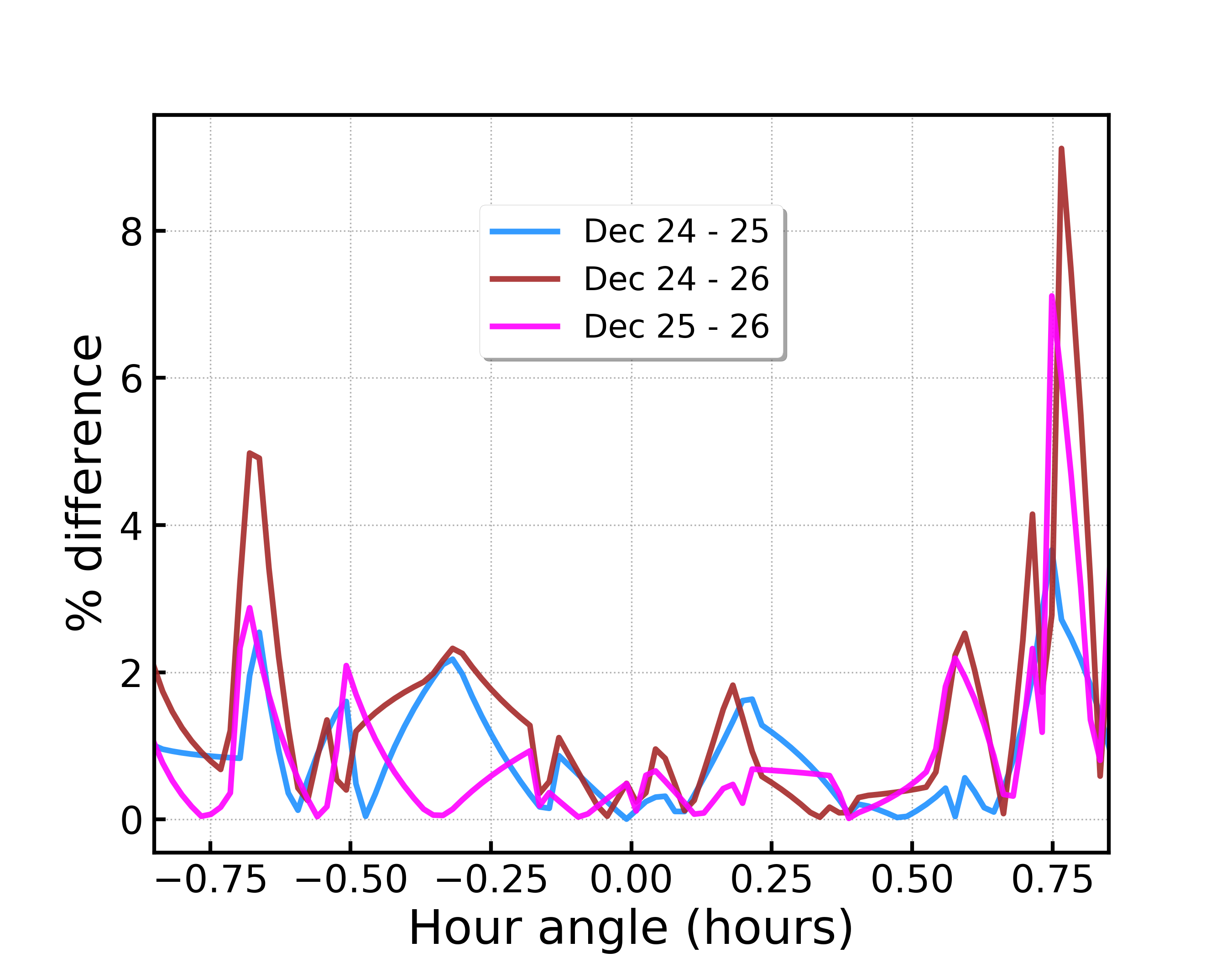}
	\caption{Difference in flux measurements between three consecutive days (December 24, 25 and 26, 2018) relative to each other. The average difference between these three consecutive days is $\sim$ 2 $\%$.}
	\label{fig:src_tracks_days}
\end{figure}

The measured flux densities of the three sources highlighted in Figure~\ref{fig:msf_images} at 150~MHz are plotted as a function of time in the left panel of Figure~\ref{fig:src_tracks}. We see that sources can be tracked down to confusion noise for about 6~hours, transiting the main lobe of the instrument for about 1.7~hours. The right panel shows the relative error associated with the measured flux values $\frac{\Delta I^{\prime}_{\nu}}{I^{\prime}_{\nu}}$ which increases as sources drift away from the main lobe due to dominance of confusion noise in our measurements. The error within the main lobe, where we assume sufficient SNR, is plotted in the top left inlet window. The error on average, represented by the dotted blue lines, is found to be less than 10$\%$.

The above analysis is conducted on 12~hours of data from December 24, 2018. In order to study day--to--day variations in the flux measurements, we repeated the aforementioned process on 12~hours of hours from December 25 and 26, 2018. The difference in flux measurements relative to these three consecutive days is shown in Figure~\ref{fig:src_tracks_days}. On average, the flux values varies by around $2\%$, but as the source moves away from zenith, we see that the difference increases up to $9\%$, consistent with the error associated with the flux extraction illustrated in Figure~\ref{fig:src_tracks}. It is worth noting that this error estimate might change if we increase the gap between days due to instrumental (e.g. system temperature) and atmospheric (e.g. ionosphere) effects.

\subsection{Beam Construction}
\label{subsec:beam_construction}
We apply the formalism derived in Section~\ref{sec:formalism} to the 113 source tracks.
The formalism is implemented in the \verb beam_solver \footnote{\url{https://github.com/Chuneeta/beam_solver}} \citep{beam_solver} Python package.
In this work, we mainly pursue the method of \citet{Pober2012} but our approach is slightly different. Firstly, since HERA has two orthogonal dipoles as its feed, we use the 90$^{\circ}$ rotation between the East--West and North--South polarizations, illustrated in bottom left plot in Figure~\ref{fig:rotated_tracks} in this work, similar to \citet{Eastwood2018} to break the degeneracy between source flux and beam response in our equations. The 90$^{\circ}$ rotation overcomes the issue of having a source track overlapping with rotated one. However, given that we construct equation~\ref{eq:corr_flux} for each data point, certain data points might repeat themselves, thus including the same equations again in the system. We therefore discard such data points to avoid duplication. By using this 90$^{\circ}$ rotation, we inevitably reduce our beam solutions to a single polarization. Secondly, we work in the spatial or image domain to obtain our flux measurements whereas \citet{Pober2012} extracted the flux density using the delay transform approach \citep{Parsons2012}. Working in the spatial domain allows to extract flux densities per frequency (discussed in Section~\ref{sec:imaging}), unlike the delay domain. Thirdly, \citet{Pober2012} derive the beam solutions with only 25 sources followed by a deconvolution process to fill in the gaps. As the FWHM of HERA is about half that of PAPER, our measurements provide an entire beam coverage.


The rationale behind the framework presented in this paper is aimed at an ongoing series to know our averaged instrument response better. For instance, the synthesized beam generated by the CLEAN algorithm does not yield the non-idealities or small deviations on the primary beam. In order to derive these non-idealities, we need a system of equations as presented in Section~\ref{sec:formalism} which includes a combination of the primary beam. Moreover, the redundantly calibrated visibilities used to derive the beam solutions were absolutely calibrated using two sources from GLEAM (refer to Section~\ref{sec:observations}), therefore the calibration solutions deliver limited information on the sources, while our beam solving algorithm uses all the sources obtained from our observations to solve for the beam and source flux densities, which in turn enables us to measure small deviations from the primary beam.

\begin{figure}[ht!]
	\centering	
	\includegraphics[width=0.95\linewidth]{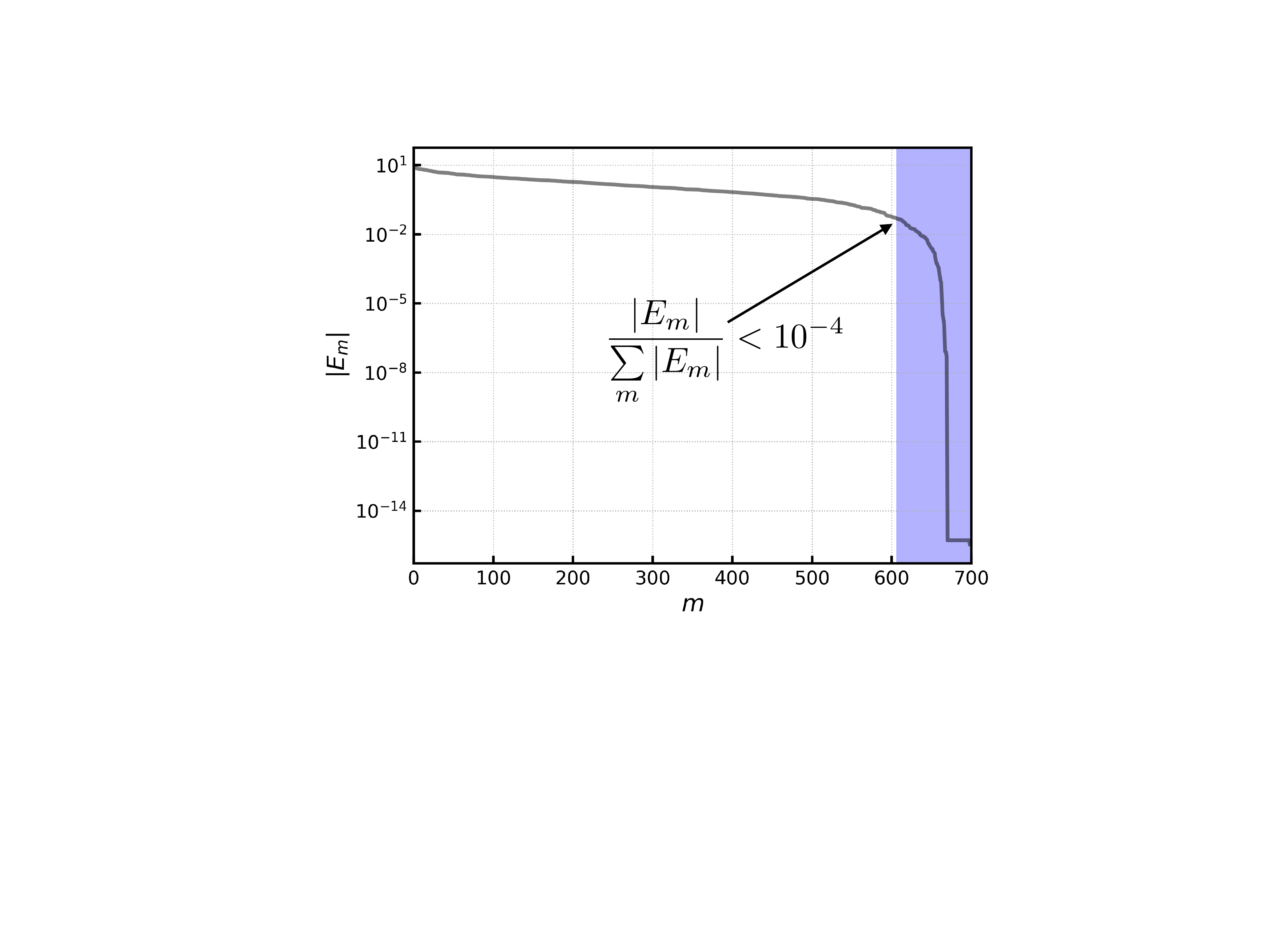}
	\caption{Eigen--amplitude, $|E_m|$ versus eigenmodes $m$ obtained by decomposing the condition ${\bf C}$. The shaded region marks the eigenmodes that are degenerate in the beam solutions.}
	\label{fig:eigenmodes}
\end{figure}

The fit to the system of equations uses the flux density values estimated from equation~\ref{eq:corr_fflux}. We do not use the EM simulations but instead give zeros as input to the primary beam, however we constrain the pixel corresponding to zenith to be unity. The beam solutions are sensitive to the grid size as mentioned in Section~\ref{subsec:least_square}, hence choosing the proper grid size is important. We choose a grid size of $1^{\circ}$ to conform the assumptions on our beam symmetries. Given that we have a datum every 10~minutes, the same data point falls into multiple grid resulting in additional degeneracies from poorly constrained pixels. We therefore interpolate between the data points every 3~minutes so that we have a unique datum for each grid. In order to estimate the uncertainties associated with the interpolated points we require to understand the correlation between our flux measurements. This derivation is analytically possible and will lead to a full correlation matrix referred to as the noise matrix $\bf {N}$ (see Section~\ref{sec:beam_measurements} for details). However, we adopt a partially correlated model instead to cut down the computational cost. This model assumes that each interpolated point is equally correlated between adjacent measurements, and as a result leads to an overestimation of the uncertainties.

Even post--interpolation, some of the pixels remain poorly constrained. We therefore decomposed the condition matrix ${\bf C}$ from equation~\ref{eq:nlinear_sys} into its corresponding eigenvalues using singular value decomposition. 
An example of the output eigen amplitudes $|E_m|$ versus eigenmodes $m$ is illustrated in Figure~\ref{fig:eigenmodes}. We notice that after eigenmode 606, there is drastic decrease in amplitude. The strange behaving eigenmodes reflect poorly constrained modes subject to numerical instability and noise, and projecting them out takes care of the degenerate modes. We discover that the amplitudes of the degenerate eigenmodes are about 4 orders of magnitude lower relative to the cumulative sum of $|E_m|$, hence we define the cut--off threshold such that all eigenmodes with $\frac{|E_m|}{\sum \limits_m{|E_m|}} < 10^{-4}$ are discarded. 

\section{Beam Pattern Measurements}
\label{sec:beam_measurements}
\subsection{Initial beam solutions}
\label{sec:initial_beamsols}
\begin{figure}[ht!]
	\centering	
	\includegraphics[width=1\linewidth]{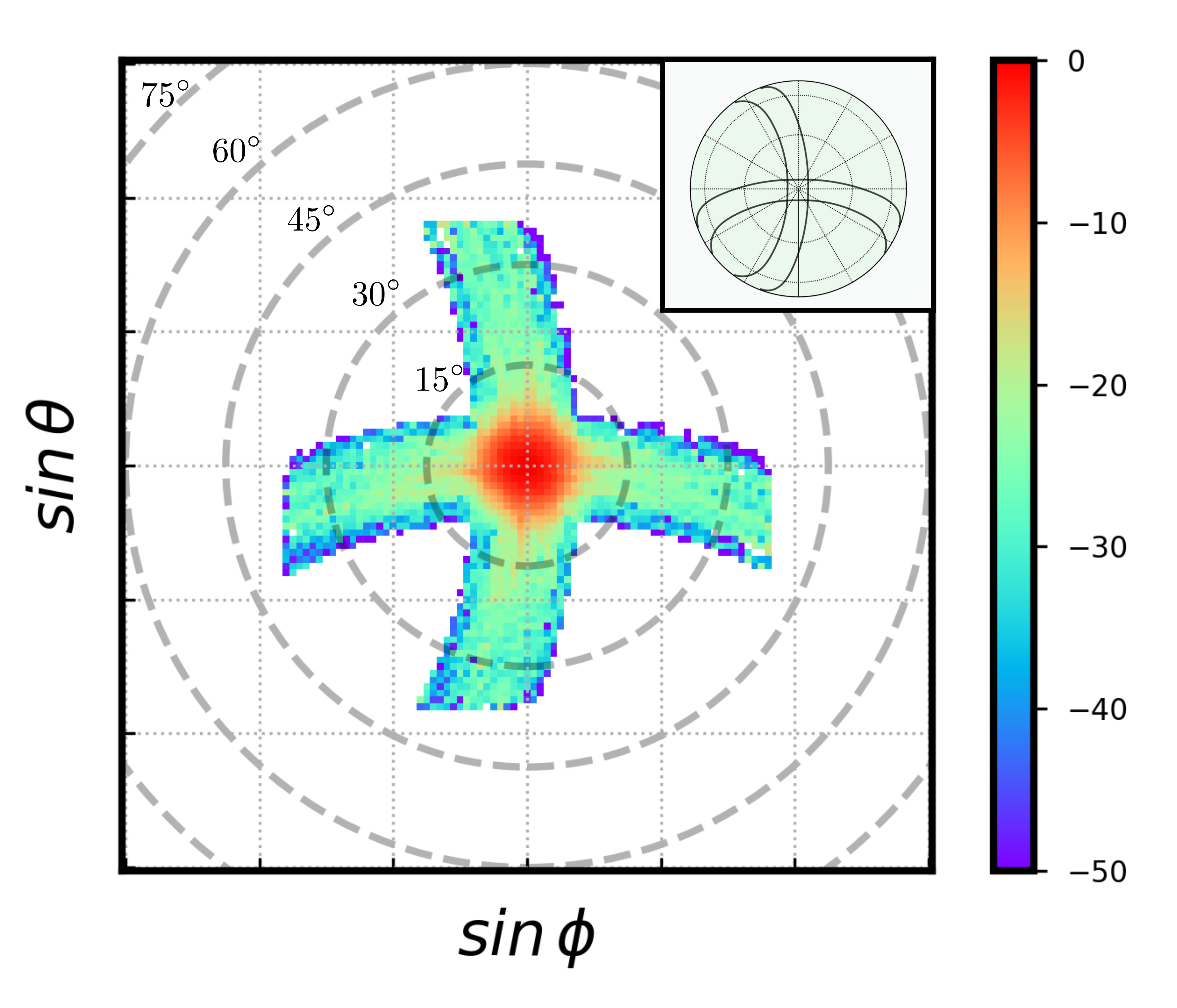}
	\caption{Beam solutions obtained using flux densities measured from images at 150~MHz. The inset shows the beam symmetry used to derive these beam solutions. The tails in the East--West and North--South directions with 10$\%$ gain amplitude relative to the peak gain throughout are suspected to be confusion noise dominated. Therefore, to probe the actual beam response,  we use measurements of known bright sources such as Pictor~A, Virgo~A,  Hercules~A, Centaurus~A and few MRT sources (see Figures~\ref{fig:strong_srctracks} and \ref{fig:faint_srctracks}).}
	\label{fig:beampattern_150MHz}
\end{figure}

The beam solutions derived at 150~MHz using the observations and following the processes described in Section~\ref{sec:observations} are shown in Figure~\ref{fig:beampattern_150MHz}. Given that the solutions are estimated from equation~\ref{eq:corr_flux}, they represent the gain amplitude of the ratio of measured flux densities to the intrinsic or expected ones. 
The gain amplitude of an antenna, measured in units of decibels (dB) refers to the energy or power transmitted in the peak direction of the radiation relative to an isotropic source and is given by 
\begin{equation}\label{eq:gain_amp}
G_{\rm{dB}} = 10 \, \rm{log}_{10} \Big(\frac{A_{\nu}}{A_0}\Big)
\end{equation}
where $A_0$ is the reference energy (of isotropic source)  and is equal to unity in our case.

The tracks formed by the beam solutions along the East--West and North--South directions demonstrate the $90^{\circ}$ beam symmetry illustrated by the inset (refer to Sections~\ref{sec:formalism} and \ref{subsec:beam_construction}). We obtain a circular--like beam unlike the EM simulations that portray an elliptical shape (Figure~\ref{fig:simulated_beam} and this is due to our measurements hitting the noise floor at or before the first null. We provide more evidences on the noise behavior in later sections.

\begin{figure}[ht!]
	\centering	
	\includegraphics[width=1.0\linewidth]{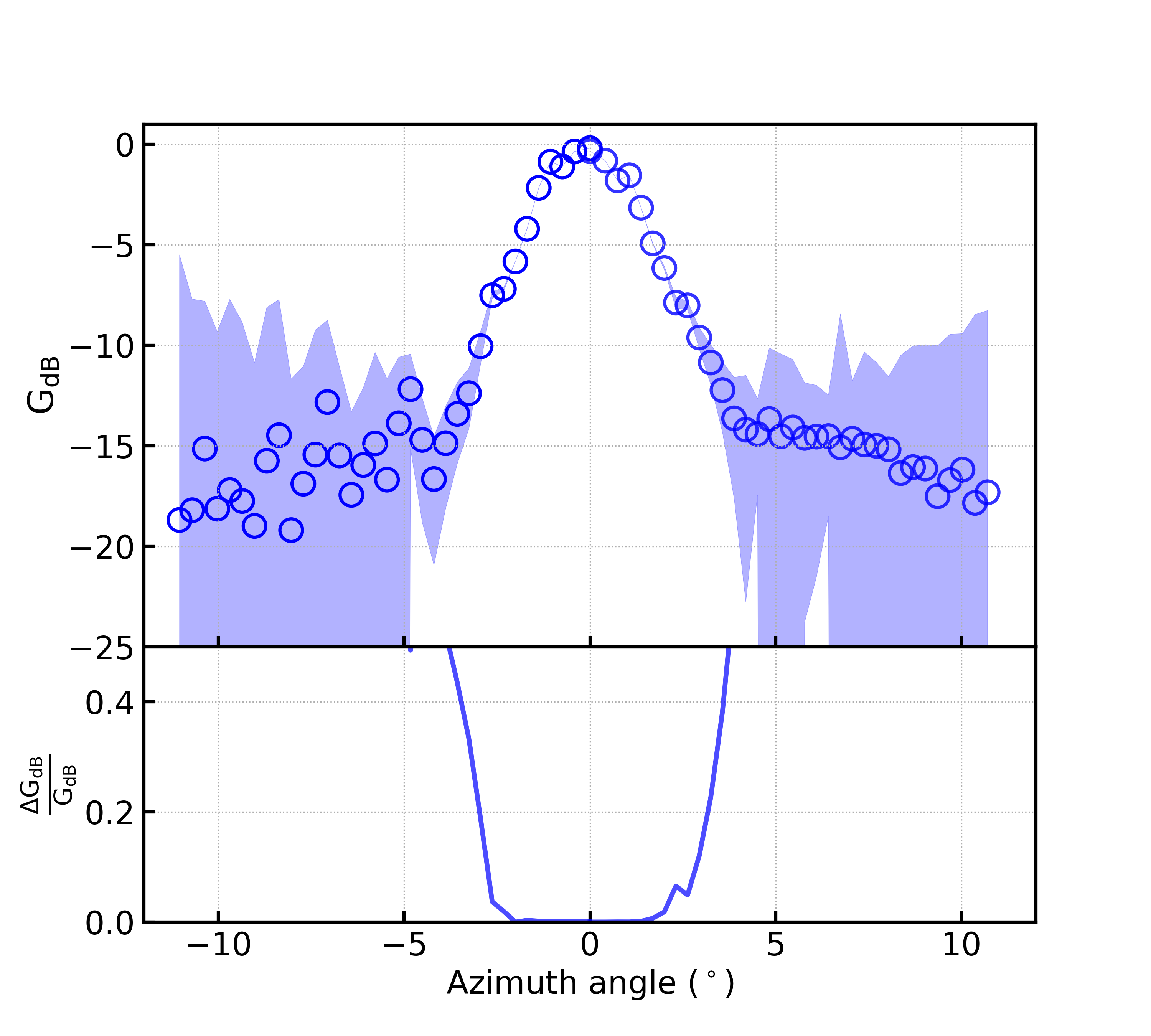}
	\caption{{\textit Top:} Beam solutions cut from Figure~\ref{fig:beampattern_150MHz} representing the gain of our instrument across the East--West polarization, crossing through zenith where it has the maximal sensitivity. Errors on the beam solutions $\Delta G_{\rm dB}$ are highlighted by the shaded region. {\textit Bottom:} Relative error with respect to the gain amplitude $\frac{\Delta G_{\rm dB}}{G_{\rm dB}}$ which is $<$5$\%$ within the main lobe and increases as it moves to the nulls.}
	\label{fig:beampattern_EW}
\end{figure}

Ideally, the beam solutions are expected to be symmetric about the center both along the East--West and North--South polarizations. We show a plot of the cut from Figure~\ref{fig:beampattern_150MHz} through zenith along the East--West polarization in the top panel of Figure~\ref{fig:beampattern_EW}. We see that the solutions within the main lobe $\sim$ 10$^{\circ}$ are almost symmetric about $\phi=$ 0$^{\circ}$, however as the azimuth angle increases to 5$^{\circ}$, the symmetry disappears. The solutions are scattered between -5$^{\circ}$ and -10$^{\circ}$ and the gain amplitudes are higher unlike those within 5$^{\circ}$ and 10$^{\circ}$. The shaded region depicts errors on the beam solutions $\Delta G_{\rm dB}$ and they are calculated using $({\bf C} {\bf N^{-1}} {\bf C^T})^{-1}$, where ${\bf N}$ is a tridiagonal matrix containing errors evaluated during the interpolation (refer to Section~\ref{subsec:beam_construction}). The relative error with respect to the gain amplitude $\frac{\Delta G_{\rm dB}}{G_{\rm dB}}$ plotted in the bottom panel is less than 5$\%$ within the main lobe and goes up to 40$\%$ in the sidelobes. 
It is also observed that the errors for $\phi <$ -7$^{\circ}$ is higher than $\phi >$ 7$^{\circ}$. The cause for this behavior may be confusion noise in our measurements.  Since sources with low SNR quickly drive down below noise level, we need measurements of few bright sources to probe the actual beam response.

\begin{figure*}[ht!]
	\centering	
	\includegraphics[width=1.0\linewidth]{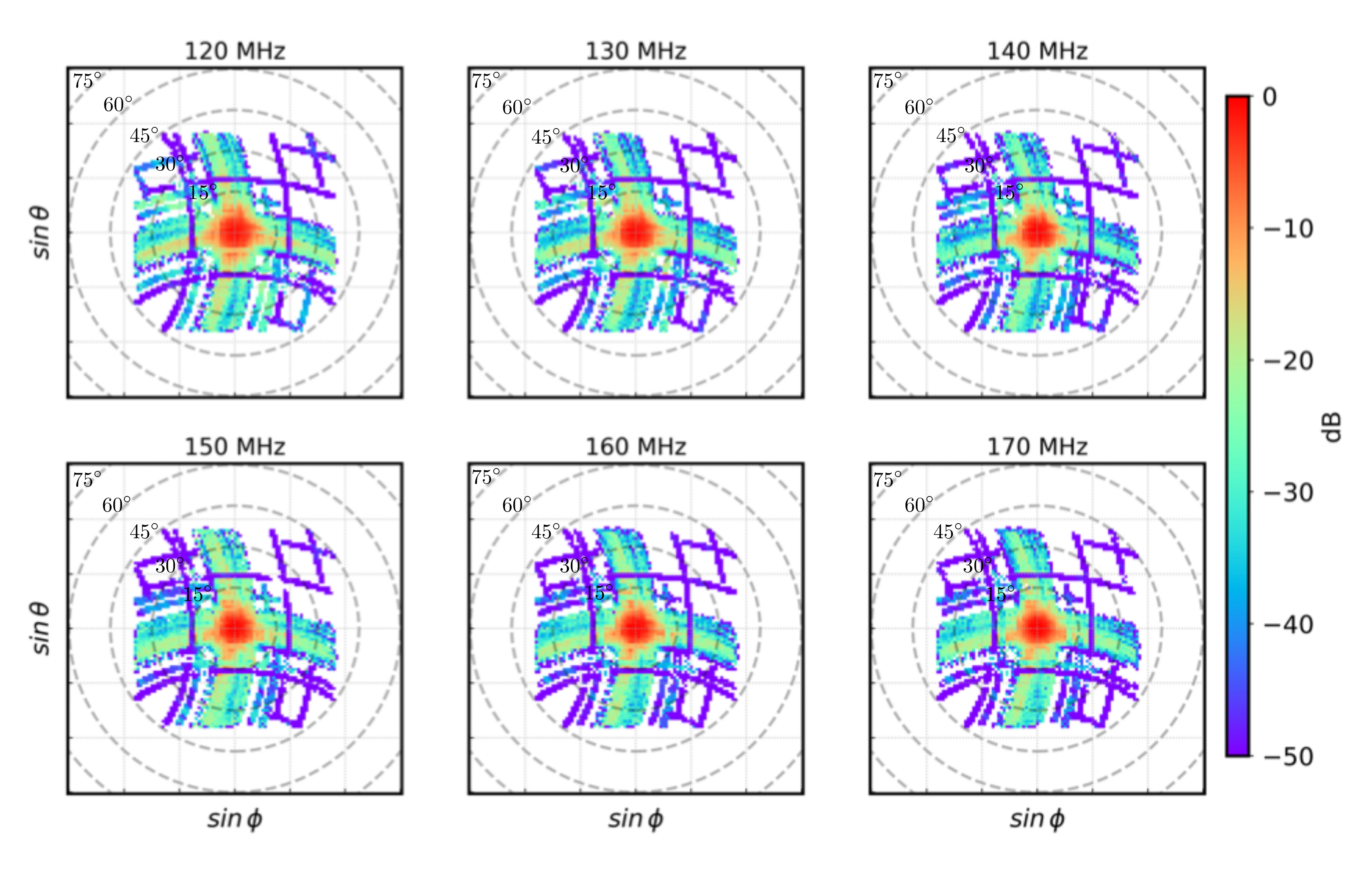}
	\caption{Beam solutions measured at various frequencies after including the bright sources. The additional tracks that we see here are the beam tracks constructed by these additional sources. The drop in amplitude confirms that the beam solutions in the tails are dominated by confusion noise.}
	\label{fig:beampattern_freq}
\end{figure*}

\subsection{Improving beam solutions}
\label{sec:improving_beamsols}
\begin{figure}[ht!]
	\centering	
	\includegraphics[width=1.0\linewidth]{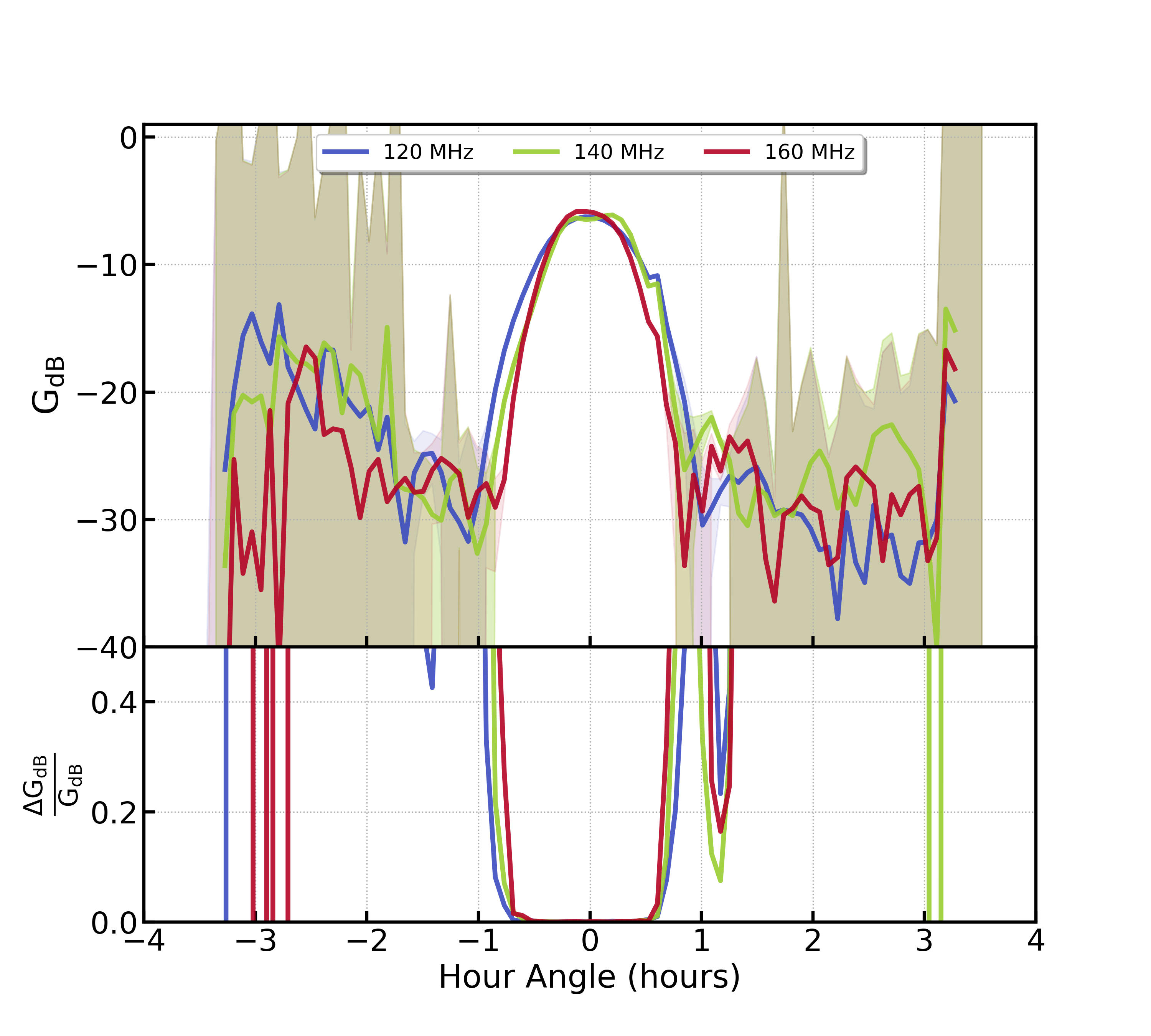}
	\caption{{\textit Top:} Regenerated tracks of Fornax~A using the derived beam solutions in Figure~\ref{fig:beampattern_freq} illustrating changing FWHM with frequency. Shaded region indicated the error on the beam solutions. {\textit Bottom:} Error associated with the beam solutions relative the gain amplitude, which shoots up to 40$\%$ beyond the main lobe.}
	\label{fig:fornaxA_freq}
\end{figure}
Bright sources such as Pictor~A, Virgo~A,  Hercules~A, Centarus~A and few of candidates from the Mauritius Radio Telescope survey \citep[MRT; ][]{Golap1998} are observed by HERA at noise level. We include these aforementioned sources in our least square optimization and constrain their corresponding intrinsic flux densities to be measurements from the literature. The flux densities of these bright sources are obtained from surveys carried out at 150~MHz \citep{Israel1998, Golap1998, Jacobs2013} at 160~MHz \cite{Slee1995}, and are scaled to the desired frequency using the power--law formula for synchrotron emission \citep{Rybicki1986}. Figure~\ref{fig:beampattern_freq} displays the improved beam solutions derived at 120, 130, 140, 150, 160, and 170~MHz, plotted on a 2-dimensional grid. The isolated tracks represent the beam solutions formed by the additional bright sources. The amplitude drop in the solutions is about an order of magnitude indicating that the beam response estimated beyond the main lobe are dominated by confusion noise in our data.

We characterize the spectral properties of the estimated beam solutions by tracing the beam response of the brightest source in our FoV, Fornax~A between hour angles of -3~hours to 3~hours. The beam tracks regenerated using the improved solutions in Figure~\ref{fig:beampattern_freq} at 120, 140 and 160~MHz are illustrated in the top panel of Figure~\ref{fig:fornaxA_freq}. The maximal sensitivity of the regenerated beam responses at the various frequencies is centered at LST$\sim$ 0~hours peaking at the same gain amplitude. We observe a decrease in FWHM with increasing frequency as the beam solid angle is inversely proportional to $\nu$ \citep{Kraus}. Similar to Figure~\ref{fig:beampattern_EW}, the beam solutions between hour angles of -3~hours and -2~hours have higher amplitude compared to those between hour angles of 2~hours and 3~hours. We speculated in Section~\ref{sec:initial_beamsols} that this phenomenon may be due to confusion noise in our measurements. However, as our beam solutions are derived using a Fourier transform implementation of the visibilities, they are basically tied to the gain solutions obtained from redundant calibration~\citep{Dillon2020} and the high SNR of Fornax~A suggests possible calibration issues. The bottom plot shows the error relative to the gain of the beam amplitude, consistent with Figure~\ref{fig:beampattern_EW}.

\subsection{Comparison with EM simulations}
\begin{figure*}[ht!]
	\centering	
	\includegraphics[width=1.1\linewidth]{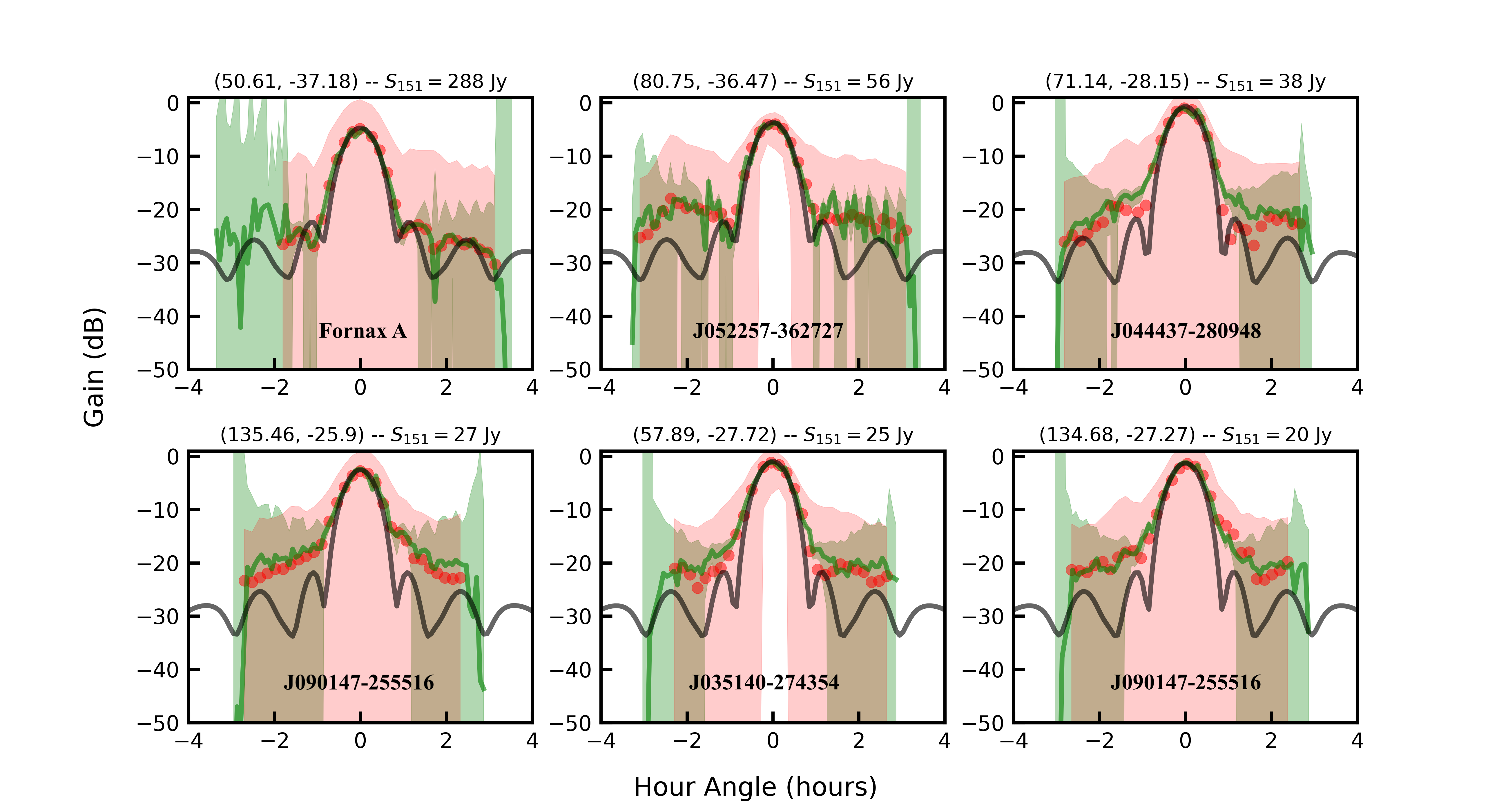}
	\caption{Gain amplitude defined in equation~\ref{eq:gain_amp} of the six brightest sources in our FoV evaluated from the beam solutions in Figure~\ref{fig:beampattern_freq} (solid green line in top panel of each subplot) and EM simulations (black) at 150~MHz. The shaded region in green denotes the errors associated with the beam solutions. The ratio of the extracted flux densities $I_{\nu}(\hat{s}, t)$ to $I_{\nu}(\hat{s})$ calculated using equation~\ref{eq:corr_fflux} are plotted in red. The shaded region in red highlights the one sigma errors on these flux measurements. The beam solutions follow our flux measurements down to the tails or sidelobes and agree with EM simulations down to -20~dB.} 
	\label{fig:strong_srctracks}
\end{figure*}

\begin{figure*}[ht!]
	\centering	
	\includegraphics[width=1.1\linewidth]{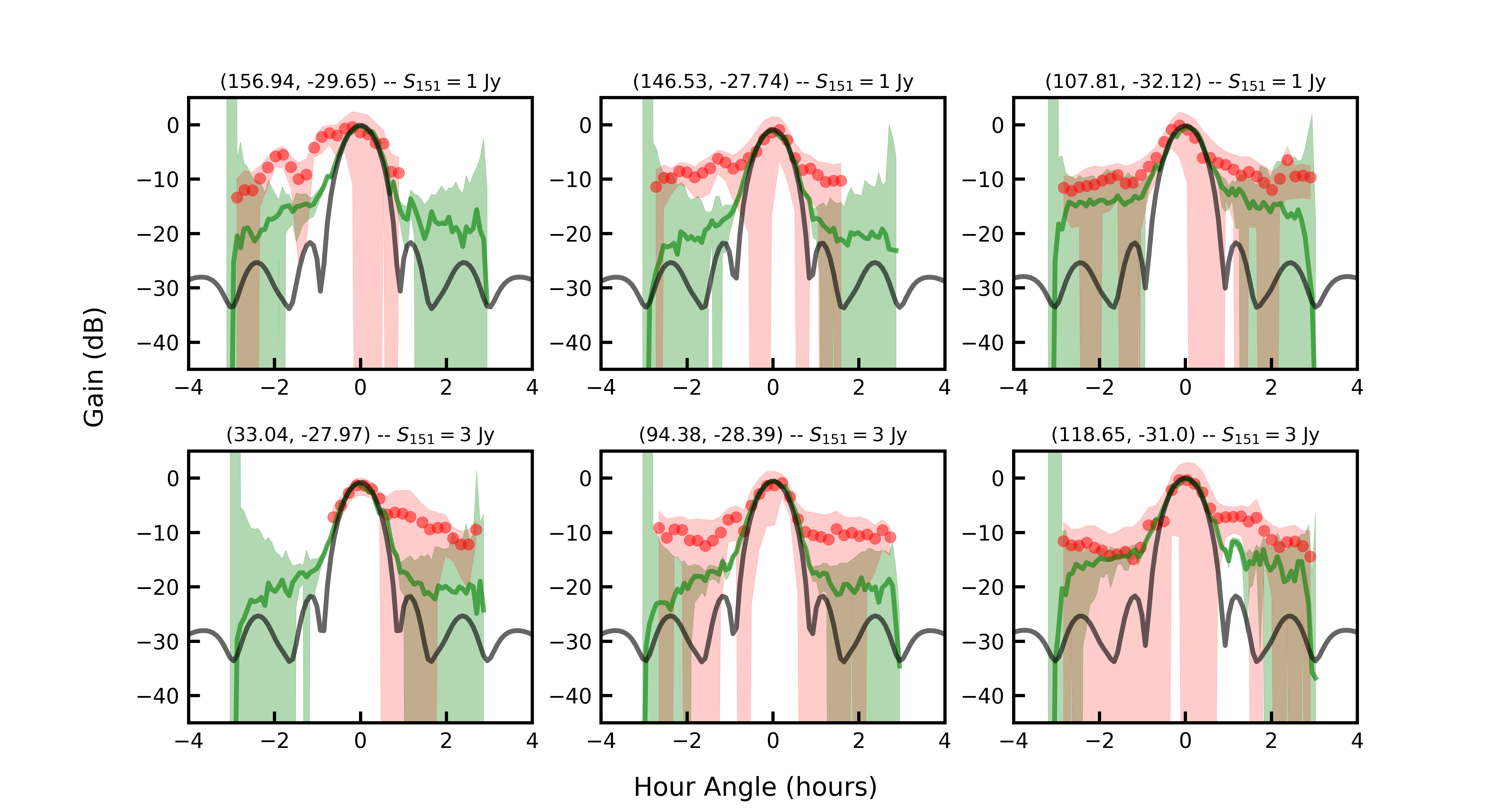}
	\caption{Same as Figure~\ref{fig:strong_srctracks} but for the six faintest sources. Here we see that for most of the sources our flux measurements are an order of magnitude higher than the beam solutions implying that the least square optimization is driven by sources with high SNR. Our beam solutions match the EM simulations in the main lobe down to -13~dB. As we move away from the main lobe, our flux measurements hit the noise floor, thus the beam solutions are likely to be confusion noise dominated.}
	\label{fig:faint_srctracks}
\end{figure*}

In this section, we compare the beam solutions at 150~MHz with the EM simulations. To begin with, we regenerated the beam responses along the East--West polarization using the improved solutions at 150~MHz for the six brightest sources in the FoV.
The regenerated beam responses are plotted in green in Figure~\ref{fig:strong_srctracks}. The position in degrees and strength in Jansky of the sources are displayed on top of each plot. 
The errors on the beam solutions $\Delta G_{\rm dM}$ are highlighted by the shaded green region.
They are consistent for all six sources and with Figure~\ref{fig:beampattern_EW}, increasing as the sources drift away from the main lobe.

The beam responses for the six brightest sources are symmetric about LST=0~hours in the main lobe, consistent with Figures~\ref{fig:beampattern_EW} and \ref{fig:beampattern_freq}. The sidelobes depict spiky features that might be fundamentally due to confusion noise as supported by the errors shaded in green, therefore they are not reliable. 
Further, we do notice the large amplitude of the beam solutions observed in Fornax~A between hour angles of $\sim$~3 and $\sim$~2~hours in the remaining five sources. The evidences seem to be pointing towards the visibilities being less redundant when Fornax~A enters the FoV \citep{Dillon2020} and this requires further investigation that is beyond the scope of this work.

We then over-plot the flux measurements in red on the beam solutions, and find that they follow the desired source tracks, which is anticipated. The flux measurements plotted in Figure~\ref{fig:strong_srctracks} are scaled such that they are equal to $I^{\prime}_{\rm{150MHz}}(\hat{s}, t)/I_{\rm{150MHz}}(\hat{s}, t)$ where $I_{\rm{150MHz}}(\hat{s}, t)$ is obtained from equation~\ref{eq:corr_fflux}. The shaded region in red shows one sigma errors associated with the flux measurements. The variations in the sidelobes of the beam solutions are consistent with the errors on our flux measurements. 
Another thing to note here is that we do not have flux measurements for Fornax~A for hour angles less than -2~hours unlike the remaining five sources and this could be a potential cause for the high amplitude of the beam solutions and, thus high error bars.

Next, we compare the beam solutions (green line) with the EM simulations plotted in black in Figure~\ref{fig:strong_srctracks}. The beam solutions matches the EM simulations down to about -20~dB relative to the peak amplitude. 
Moreover, with high SNR sources such as Fornax~A, we could track the beam responses down to the first null as observed in the top left panel of Figure~\ref{fig:strong_srctracks}. At the same time, we notice that the beam solutions beyond the main lobe for the other sources are about an order of magnitude higher than the EM simulations that can be attributed to increasing confusion noise in the sidelobes. 

We repeat the aforementioned comparison for the six faintest sources and the plots are illustrated in Figure~\ref{fig:faint_srctracks}. In this case, our measurements are able to go down to -10~dB only while the beam solutions in the main lobe goes down to -20~dB for most of the sources, apart from those displayed in the right panel. Beam solutions that go down to -20~dB seem to have fewer flux measurements (they are not fully tracked by our observations) as compared to those that go down to -10~dB, indicating that sources with high SNR are advantageous in improving our beam solutions. 
The discrepancies in our beam solutions  are consistent for all the sources within two sigma. Note that the errors on the flux measurements (region shaded in red) shows the errors to one sigma.

The beam solutions in the main lobe agree down to about -13~dB level relative to the peak gain with the EM solutions for the sources below 3~Jy. Similar to the bright sources, the beam solutions in the sidelobes are an order of magnitude higher than the EM simulations for the non--fully tracked sources and are about two orders of magnitude brighter for the fully tracked sources.  The measurements involved in deriving the beam solutions in the sidelobes are mostly confusion noise, therefore we expect the solutions to be highly contaminated by confusion noise.

We then rerun the least square optimization, but this time excluding the 22 sources that are located within the Magenallic Clouds and excluded in the GLEAM survey (refer to Section~\ref{sec:source_extraction}) to study the impact of these sources on our beam solutions. The resulting beam solutions do not show any variations for bright sources, however slight distortions are seen in the fainter ones where confusion noise is the limiting factor. 
Since the brightest source among these excluded sources is about 31~Jy and the remaining ones are less than 15~Jy, we confirm that that the beam solutions are sensitive to high SNR sources.

\begin{figure}[ht!]
	\centering
	\includegraphics[width=1.01\linewidth]{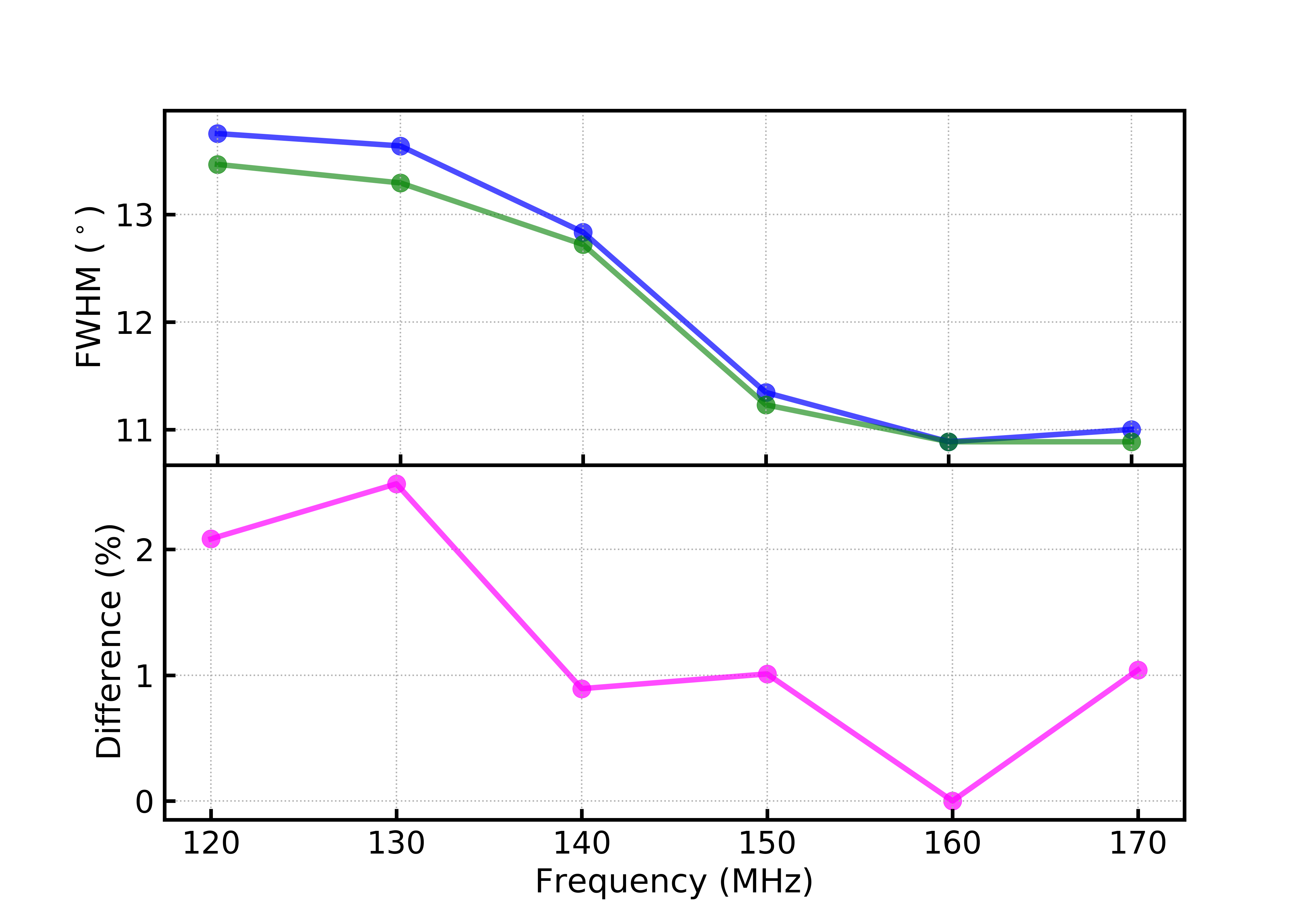}
	\caption{{\textit Top}: FWHM evaluated using the beam solutions along the East--West polarization illustrated in Figure~\ref{fig:beampattern_freq} (blue) and EM simulations (green) at various frequencies. {\textit Bottom}: The percentage difference in FWHM relative to the EM simulations. We observe a good agreement between the beam solutions and EM simulations.}
	\label{fig:fwhm_freq}
\end{figure}

To further validate the beam solutions, we compare the FWHM of the beam solutions along the East--West polarization evaluated from Figure~\ref{fig:beampattern_freq} with the EM simulations. The FWHM estimated for an arbitrary source transiting zenith and the percentage difference relative to the EM simulations as a function of frequency are plotted in the top and bottom panels of Figure~\ref{fig:fwhm_freq} respectively. We clearly see the decrease in FWHM with frequency as stated in Section~\ref{sec:improving_beamsols} and, our beam solutions agree with the EM simulations within less than $3\%$.

\section{Source Catalogue}
\label{sec:source_cat}
We also estimate flux densities for the 113 sources that went into the least square optimization. We construct a source catalog displayed in Table~\ref{tab:hera_srcs} using these flux estimates. It covers an area of $\sim$3600 deg$^2$. We then compare the estimated flux densities with the GLEAM survey \citep{Hurley-Walker2017} and find counterparts for 90 sources (refer to Figure~\ref{fig:hera_gleam_pos}).  Sources in the GLEAM survey are resolved at an angular resolution of $2^{\prime}$ while our source measurements are resolved at 
$8^{\prime}$, revealing potential candidates being unresolved by HERA unlike GLEAM measurements. For such sources, we add all the resolved GLEAM sources found within HERA's synthesized beam to match the flux density measured by HERA and use the position of the brightest source in the cluster to match the position of our sources. The measurements of these matched sources are reported in Table~\ref{tab:hera_srcs}. 

A description of the source catalog is as follows:
\begin{itemize}
	\setlength\itemsep{0.1em}
	\item {\bf Column 1}: Source ID, taken from GLEAM survey. If the flux density reported is the sum of multiple sources, $*$ is added at the end of the source ID corresponding to the brightest one in the cluster;
	\item {\bf Column 2}: Right ascension $(\alpha)$ in degrees of the position measured by PyBDSF;
	\item {\bf Column 3}: Right ascension $(\alpha_0)$ in degrees of the matched counterpart from the GLEAM survey;
	\item {\bf Column 4}: Declination $(\delta)$ in degrees of the position measured by PyBDSF;
	\item {\bf Column 5}: Declination $(\delta_0)$ in degrees of the matched counterpart from the GLEAM survey;
	\item {\bf Column 6}: Estimated flux density at 151~MHz in Jy $S_{151}$;
	\item {\bf Column 7}: Flux density of the matched source(s) measured by the GLEAM survey at 151~MHz  $S_{G151}$.
\end{itemize}

\begin{figure}[ht!]
	\centering	
	\includegraphics[width=1.\linewidth]{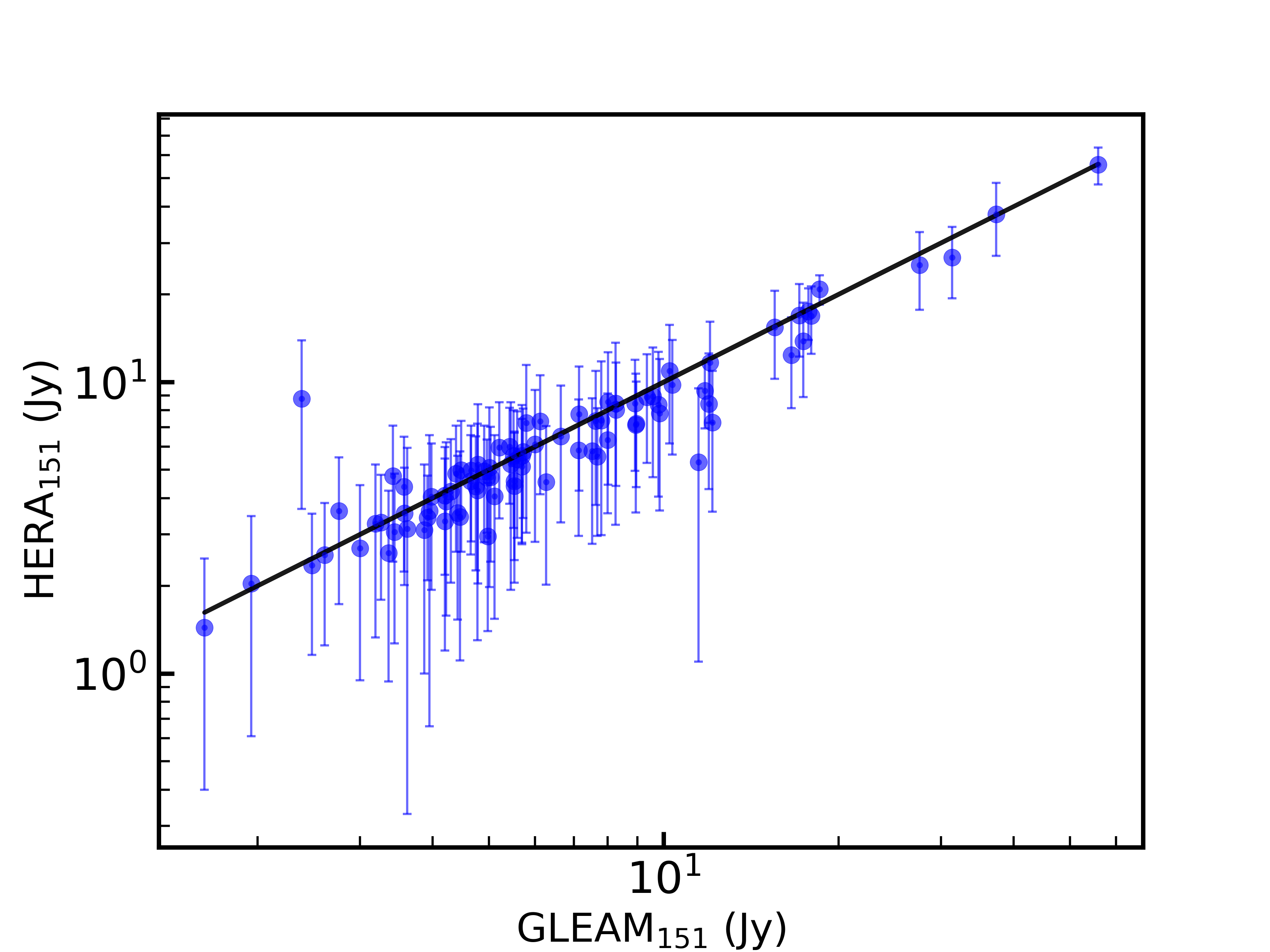}
	\caption{Flux densities of the 90 matching components from GLEAM $S_{G151}$ plotted against the derived flux densities $S_{151}$ reported in Table~\ref{tab:hera_srcs}. Most of our flux measurement align with the GLEAM results within the quoted errors except a few namely `J061721-282547', `J094953-251138', `J103312-341842'.}
	\label{fig:hera_gleam}
\end{figure}

We examine the positional difference between our measurements and matched GLEAM counterpart, plotted in Figure~\ref{fig:hera_gleam_pos} and find that sources that expose an absolute positional difference of about $30\%$ are resolved by the GLEAM survey. Since ($\alpha_0$, $\delta_0$) for such sources correspond to the position of the brightest source in the cluster, slight phase shifts can be expected.

\begin{figure}[ht!]
	\centering	
	\includegraphics[width=1.2\linewidth]{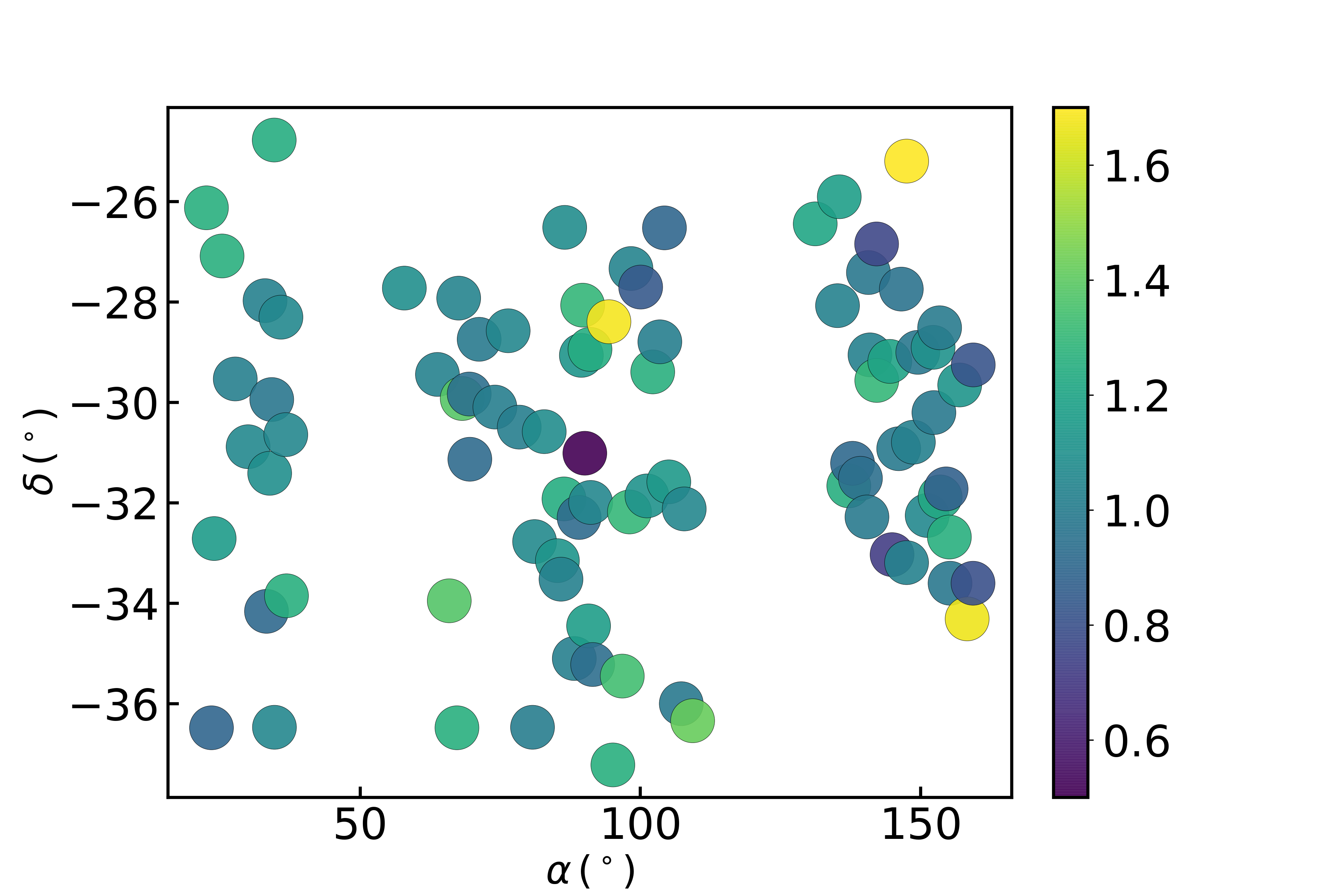}
	\caption{Scatter plot of the right ascension vs declination of the matching sources to examine trends in our flux densities versus source locations. The color of each source indicates the ratio of the derived flux densities to that of matching GLEAM component. We found no correlation between the flux densities and the source locations. The three discrepant sources seen in Figure~\ref{fig:hera_gleam} depict ratios $>$ 1.5.}
	\label{fig:radec}
\end{figure}

Figure~\ref{fig:hera_gleam} shows the flux densities of the GLEAM counterparts versus our flux density estimates. Our results are consistent with the GLEAM measurements within the estimated errors, except a few sources (`J061721-282547', `J094953-251138', `J103312-341842'). To further study the discrepant behavior of these sources, we plotted the ratio of our flux density estimates to GLEAM measurements. We observe no obvious correlation with the ratios and source positions. The ratio for these sources go up to 1.5, therefore difference in telescope resolution cannot be blamed. This discrepant behavior may be due to errors involved in the calibration process that are more pronounced in off-zenith sources.

\section{Effects of confusion noise}
\label{sec:confusion_noise}
In this section, we investigate the effects of confusion noise on our results and, test the robustness of our method.
For each time sample and frequency channel, we perform the following operations:
\begin{enumerate}
	\item generate the sky emission evaluating equation~\ref{eq:corr_flux} where the sky model consists of a catalog of point sources. The primary beam response $A_{\nu}$ is calculated from EM simulations;
	\item simulate visibilities via a discrete Fourier Transform \cite{Thompson2017} using the tools available in CASA;
	\item form images and extract the flux densities of sources as per Sections~\ref{sec:imaging}  and \ref {sec:source_extraction} respectively;
	\item construct a non linear least square problem (Section~\ref{sec:formalism}) and solve for the beam solutions and flux densities as described in Section~\ref{subsec:beam_construction};	
	\item repeat steps (1) to (4) adding Gaussian noise to the sky emission. The RMS is taken to be about 10$\%$ the peak amplitude of the source.

\end{enumerate}

We first use a single 1~Jy point source catalog and carry out the simulations. We are able to fully recover the input flux density and beam responses in both the noisy and noiseless cases depicting that thermal noise has no significant effect on our results. With a single source we have no confusion noise and therefore validates our formalism.

\begin{figure}[ht!]
	\centering
	\includegraphics[width=0.5\textwidth]{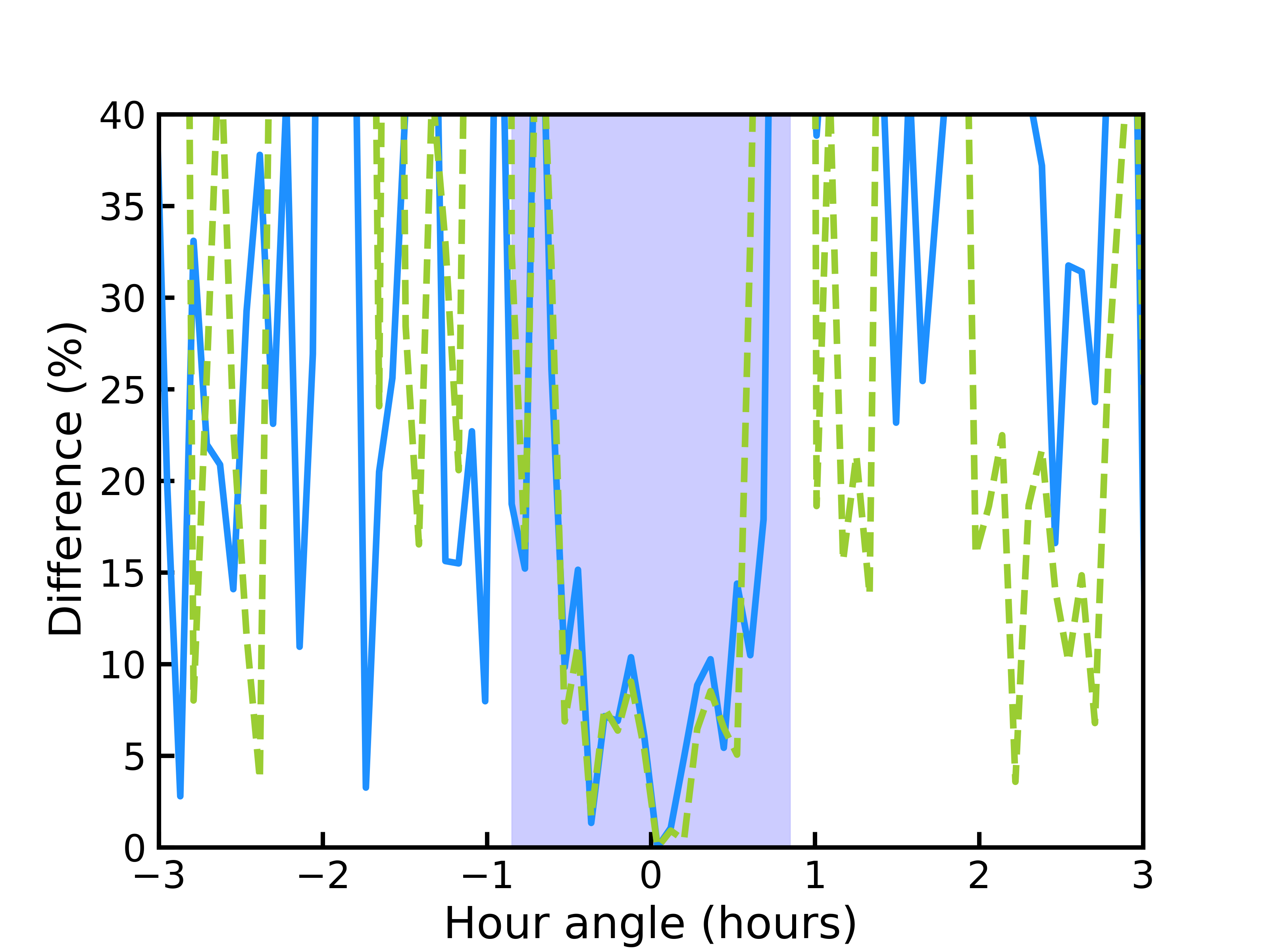}
	\caption{Simulations demonstrating that confusion noise is the dominant systematics in our approach. Adding Gaussian noise to our simulations yields negligible change inside the main lobe. We plot the percentage difference
		between the input and recovered beam of an arbitrary source transiting evaluated from a noiseless simulation in blue. We added Gaussian noise to the simulations and the resulting difference is plotted in green. The shaded area represents the main lobe of the beam response.}
	\label{fig:constructed_track_sim_1src}
\end{figure}

The simulation is then extended to the source catalog comprising 113 sources derived in Section~\ref{sec:source_cat}. We regenerate the source tracks for an arbitrary source from the beam solutions obtained using the noiseless simulations. We calculate the difference between the EM simulations and the regenerated source tracks, illustrated in blue in Figure~\ref{fig:constructed_track_sim_1src}.  The errors in the simulated beam solutions in the main lobe (shaded region) are $<$15$\%$ and increases up to 40$\%$ in the sidelobes. This behavior is consistent with our measurements shown in Figures~\ref{fig:strong_srctracks} and \ref{fig:faint_srctracks}. The errors highlighted here indicates a combination of the errors associated with the flux extraction (Figure~\ref{fig:src_tracks}) and the errors in the beam formalism (Figure~\ref{fig:fwhm_freq}).

Including Gaussian noise to the sky emission do not report major changes (green; Figure~\ref{fig:constructed_track_sim_1src}) within the main lobe. Nevertheless, the difference varies outside the main lobe. The difference seem to be smaller for some hour angles and higher for others compared to the noiseless case. Since the confusion noise becomes significant beyond the main lobe, the average between thermal noise and confusion noise may trigger these fluctuations.

\section{Conclusions}
\label{sec:conclusions}
We have presented a technique to measure the empirical beam pattern using widefield drift--scan observations from HERA. The technique is driven by the ability of our instrument to track the flux density of sources in the FoV as they pass overhead. We formed snapshot images from a 10~MHz band centered at 120, 130, 140, 150, 160 and 170~MHz using 10~minutes observations from HERA Phase I, exposing 113 sources in the FoV. We measured the flux densities of these sources from the images and built a non linear least square problem. We formed a network of overlapping sources using a 90$^\circ$ rotation about the East--West polarization (lower right panel of Figure~\ref{fig:rotated_tracks}) to represent the beam response in the North--South polarization. This beam symmetry breaks the degeneracy between flux density and beam solution in the least square optimization.

The aim of the formalism derived in this paper is to measure small deviations present in our primary beam without the need of precise knowledge of the sky. It uses a least square approximation that allows to solve for the primary beam and flux values simultaneously. The system require initial inputs to the flux densities and beam solutions to converge.

The amplitudes of our initial beam solutions in the sidelobes are about 10$\%$ relative to the peak in the main lobe. We therefore included some known bright sources observed at noise level by HERA to probe the actual beam response in our least square optimization. The beam tracks formed by these additional sources reveals confusion noise in the sidelobes affirming that we should not rely on the beam solutions beyond the main lobe.

We further used our beam solutions to generate the beam responses of the six brightest and faintest sources in the FoV.
Our beam estimates match the EM simulations in the main lobe down to a -20~dB and -~13~dB level relative to the peak gain for sources with high and low SNR respectively. They are 1--2 orders of magnitude higher in the sidelobes. This behavior is caused by confusion noise in our flux measurements and therefore can not be trusted as stated in the paragraph above. Our results demonstrate that the elements in the array are well-illuminated by the feeds in the main lobe with sensitivity levels consistent with the EM simulations.

When evaluating the 21~cm H1 power spectrum \citep{Parsons2012b}, the power is required to be normalized by the power squared beam of the observing instrument (see Appendix of \citep{Parsons2012b}). Since our beam solutions are less noise dominated for high SNR sources, we made a conservative estimate of this power square beam using the beam solutions for the six brightest sources. When compared with EM simulations, the difference is about $\sim$ 5$\%$ that may be significant enough to contaminate the cosmological signal. However, our measurements are prone to confusion noise and calibration errors, therefore improving on our instrument's sensitivity and calibration methodologies should improve our beam measurements.

We also generated a catalog of 113 sources at 151~MHz covering an area of $\sim$3600 deg$^2$ from the flux densities estimated in the least square optimization. The flux density estimates agree with the measurements from the GLEAM survey \citep{Hurley-Walker2017} within 10--15$\%$, except a few sources. These discrepancies might be due to calibration errors, however further investigation is required that is beyond the sphere of this work.

Additionally, we carried out simulations to test the accuracy of our technique and at the same time study the impact of confusion noise on our observations. We used the EM simulations as our primary beam response to the sky emission. The beam solutions seem to agree with the EM solutions within $<$15$\%$ within the main lobe. This error incorporates the error involved in the extraction process as well, which is about $10\%$, indicating that our least square approximate implementation is accurate down to about $5\%$. We also found that thermal noise has negligible influence on our beam solutions inside the main lobe.

This work can be extended to include mutual coupling between neighboring antennas. We excluded this effect in this paper as our observations were calibrated using EM simulations excluding mutual coupling. Moreover, we assumed that all elements in the array have the same beam response throughout this work which does not hold in practice. The variations in beam responses from antenna--to--antenna contribute to modeling errors in calibration, particularly for widefield observations and this may result in significant contaminations to the cosmological signal.

\section{Acknowledgments}
This  material is  based upon work supported by the National Science Foundation under Grant Nos.1636646 and 1836019, and  institutional support from the HERA collaboration partners. This research is funded in part by the Gordon and Betty Moore Foundation. Parts of this research were supported by the Australian Research Council Centre of Excellence for All Sky Astrophysics in 3 Dimensions (ASTRO 3D), through project number CE170100013. HERA is hosted by the South African Radio Astronomy Observatory, which is a facility of the National Research Foundation, an agency of the Department of Science and Innovation. G.B. acknowledges funding from the INAF PRIN-SKA 2017 project 1.05.01.88.04 (FORE--CaST), support from the Ministero degli Affari Esteri della Cooperazione Internazionale-Direzione Generale per la Promozione del Sistema Paese Progetto di Grande Rilevanza ZA18GR02 and the National Research Foun- dation of South Africa (Grant Number 113121) as part of the ISARP RADIOSKY2020 Joint Research Scheme, from the Royal Society and the Newton Fund under grant NA150184 and from the National Research Foundation of South Africa (grant No. 103424). 

\bibliographystyle{aasjournal}
\nocite{*}
\bibliography{Bibliography}

\begin{table*}
	\caption{List of sources used in this work and their derived flux densities.}
	\label{tab:hera_srcs}
	\centering
	\begin{tabular}{ccccccc}
		GLEAM ID & $\alpha (^\circ)$ & $\alpha_0 (^\circ)$ & $\delta (^\circ)$ & $\delta_0 (^\circ)$ & $S_{151}$ (Jy) & $S_{G151}$ (Jy)\\
		\hline \hline
	J013027-260956 & 22.59 & 22.62 & -26.12 & -26.17 & 9.34 $\pm$ 2.40 & 11.76\\
	J013411-362913 & 23.49 & 23.55 & -36.48 & -36.49 & 20.84 $\pm$ 2.42& 18.55\\
	J013527-324135* & 23.97 & 23.86 & -32.71 & -32.69 & 3.14 $\pm$ 2.81 & 3.62\\
	J015035-293158* &  27.73 & 27.65 & -29.53 & -29.53 & 17.45 $\pm$  3.52 & 17.75 \\
    J014127-270606 & 25.38 & 25.36 & -27.08 & -27.17 & 6.34 $\pm$ 2.79 & 8.0\\
    J020012-305324 &  30.02 & 30.05 & -30.88 & -30.89 & 16.89 $\pm$ 4.41 & 17.95 \\  
     J021209-280009  & 33.04 & 33.04 & -27.97 & -28.0 & 2.55 $\pm $ 1.30& 2.61\\
    J021317-341325* & 33.3 & 33.32 & -34.16 & -34.22 & 6.0 $\pm $ 2.17 & 5.43\\ 
    J021527-312144 & 33.87 & 33.86 & -31.41 & -31.36 & 3.62 $\pm $ 2.96 & 3.95\\
    J021736-294750* & 34.24 & 34.4 & -29.94 & -29.8 & 3.27 $\pm $ 1.94& 3.19\\
     J021843-244814 & 34.68 & 34.68 & -24.77 & -24.8 & 7.2 $\pm $ 2.84& 8.97\\
     J021902-362603 & 34.72 & 34.76 & -36.47 & -36.43 & 8.87 $\pm$ 3.59 & 9.36\\   
   J022343-281856 & 35.88 & 35.93 & -28.3 & -28.32 & 7.38 $\pm$ 4.39& 7.81\\
   J022720-303746 & 36.75 & 36.83 & -30.64 & -30.63 & 4.7 $\pm $ 1.66  & 4.96\\
   J022716-335232 & 36.87 & 36.82 & -33.85 & -33.88 & 4.06 $\pm$ 2.52 & 5.11\\
   J035140-274354* & 57.89 & 57.92 & -27.72 & -27.73 & 25.19 $\pm$ 7.51& 27.55\\
   J041508-292901 & 63.79 & 63.79 & -29.44 & -29.48 & 8.04 $\pm$ 3.64& 8.27\\
   J042347-340234 & 65.9 & 65.95 & -33.95 & -34.04 & 4.54 $\pm$ 2.52& 6.27\\
    J042940-363050 & 67.27 & 67.42 & -36.48 & -36.51 & 13.81 $\pm$ 4.94& 17.39\\
   J043018-280045 & 67.58 & 67.57 & -27.92 & -28.01 & 5.23 $\pm$ 3.29& 5.45\\
   J043300-295609 & 68.21 & 68.25 & -29.92 & -29.94 & 5.55 $\pm$ 2.58& 7.68\\
   J043736-295359* & 69.46 & 69.4 & -29.83 & -29.9 & 8.55 $\pm$ 4.10& 8.02\\
   J043832-311243* & 69.57 & 69.64 & -31.13 & -31.21 & 5.22 $\pm$ 3.18 & 4.79\\
   J044437-280948 &  71.23 & 71.16 & -28.74 & -28.16 & 37.7 $\pm$ 10.6 & 37.34\\
   J045514-300646 &  74.04 & 73.81 & -30.09 & -30.11 & 16.95 $\pm$ 4.73 & 17.12\\
   J050519-281628* & 76.41 & 76.33 & -28.57 & -28.27 & 9.79 $\pm$ 4.15& 10.34\\
   J051135-301119* & 78.38 & 77.9 & -30.49 & -30.19 & 15.42 $\pm $ 5.17& 15.53\\
   J052257-362727 & 80.75 & 80.74 & -36.47 & -36.46 & 55.7 $\pm$ 8.04& 55.94\\
   J052333-325119* & 81.12 & 80.89 & -32.77 &-32.86 & 8.93 $\pm$ 4.21 & 9.57\\
   J053115-303210 & 82.84 & 82.82 & -30.58 & -30.54 & 3.9 $\pm$ 2.32& 4.22\\
   J054017-330918 & 85.19 & 85.07 & -33.15 & -33.16 & 4.25 $\pm$ 2.95 & 4.78\\  
   J054358-333629*& 85.83 & 85.99 & -33.52 & -33.61 & 4.21 $\pm$ 4.16& 4.3\\
   J054516-315853 & 86.36 & 86.32 & -31.92 & -31.98 & 3.56 $\pm$ 2.03& 4.42\\
   J054558-263015 & 86.49 & 86.49 & -26.51 & -26.5 & 4.39 $\pm$ 2.13& 4.75\\
   J055616-322310 & 87.57 & 89.07 & -31.65 & -32.39 & 7.1 $\pm$ 5.35& 7.15\\
   J055205-345955*& 88.2 & 88.02 & -35.1 & -35.0 & 4.57 $\pm$ 2.01& 4.65\\
   J055616-322310 & 89.06 & 89.07 & -32.29 & -32.39 & 7.77 $\pm$ 3.53& 7.15\\
   J055759-285546 & 89.46 & 89.5 & -29.06 & -28.93 & 3.06 $\pm$ 1.79& 3.44\\
   J055820-280912 & 89.69 & 89.59 & -28.06 & -28.15 & 3.45 $\pm$ 2.34& 4.46\\
   J060015-305632*&  90.08 & 90.06 & -31.01 & -30.94 & 8.78 $\pm$ 5.11& 2.38\\
   J060312-342632 & 90.75 & 90.8 & -34.45 & -34.44 & 8.38 $\pm$ 4.33& 9.79  \\
    J060414-315555 & 91.12 & 91.06 & -31.99 & -31.93 & 8.45 $\pm$ 3.49& 8.92\\
   J060405-285843  & 91.0 & 91.02 & -28.94 & -28.98 & 3.11 $\pm$ 2.11& 3.87\\
   J060554-351806* & 91.48 & 91.48 & -35.22 & -35.3 & 10.94 $\pm$ 4.78& 10.23 \\
    J061721-282547  & 94.38 & 94.34 & -28.39 & -28.43 & 2.96 $\pm$ 1.56& 4.98\\
   J062000-371133 & 95.08 & 95.0 & -37.22 & -37.19 & 7.82 $\pm$ 4.19& 9.83
   	\end{tabular}
\end{table*}

   \begin{table*}
   	\centering
   	\begin{tabular}{ccccccc}
   GLEAM ID & $\alpha (^\circ)$ & $\alpha_0  (^\circ)$ & $\delta  (^\circ)$ & $\delta_0  (^\circ)$ & $S_{151} $ (Jy) & $S_{G151} $ (Jy)\\
   		\hline \hline
    J062707-352908 & 96.76 & 96.78 & -35.45 & -35.49 & 12.39 $\pm$ 4.26 & 16.58\\
   	J063433-271116 & 98.31 & 98.64 & -27.33 & -27.19 & 7.35 $\pm$ 3.56& 7.64\\
   J063149-321654 & 98.04 & 97.96 & -32.18 & -32.28 & 2.59 $\pm$ 1.65& 3.36\\
   J063957-274532 & 100.04 & 99.99 & -27.7 & -27.76 & 7.33 $\pm$ 3.21& 6.13\\
    J064413-315552 & 101.16 & 101.06 & -31.86 & -31.93 & 2.69 $\pm$ 1.74 & 3.0\\
    J064925-291919* & 102.19 & 102.35 & -29.39 & -29.32  & 3.33  $\pm$ 2.13 & 4.2\\
   J065419-285218* & 103.46 & 103.58 & -28.79 & -28.87 & 3.55  $\pm$ 1.54 &3.58\\
   J065818-263239* & 104.28 & 104.58 & -26.52 & -26.54 & 4.99 $\pm$ 2.37 & 4.48\\
   J065716-320328 & 105.06 & 104.32 & -31.58 & -32.06 & 3.43 $\pm$ 1.34 & 3.92\\
   J071143-320300* & 107.81 & 107.93 & -32.12 & -32.05 & 2.35 $\pm$ 1.19& 2.48\\
   J070901-355921* & 107.28 & 107.26 & -36.0 & -35.99 & 5.59 $\pm$ 2.43& 5.51\\
   J071706-362140 & 109.31 & 109.28 & -36.34 & -36.36 & 8.42 $\pm$ 4.13 & 11.95\\
   J084456-263332*& 131.17 & 131.23 & -26.44 & -26.56 & 4.57 $\pm$ 2.12& 5.53\\
   J090147-255516 & 134.68 & 135.45 & -27.27 & -25.92 & 19.6 $\pm$ 4.22& 31.35\\
   J090015-281756* & 135.15 & 135.06 &-28.07 & -28.3 & 5.43 $\pm$ 2.51& 5.59\\
   J090147-255516 & 135.46 & 135.45 & -25.9 & -25.92 & 26.72 $\pm$ 7.31& 31.35\\
   J090911-313334* & 137.16 & 137.3 & -31.66 & -31.56 & 4.41 $\pm$ 2.36& 5.53\\
   J091123-310736* & 137.83 & 137.85 & -31.21 & -31.13 & 4.85 $\pm$ 2.23& 4.39\\
   J091655-313637* & 139.24 & 139.23 & -31.51 & -31.61 & 4.97 $\pm$ 2.13& 4.66\\
   J092121-321935* & 140.4 & 140.34 & -32.28 & -32.33 & 4.95 $\pm$ 2.13& 4.91\\
   J092252-272627* & 140.66 & 140.72 & -27.41 & -27.44 & 4.05 $\pm$ 2.11& 3.98\\
   J092410-290602* & 140.94 & 141.04 & -29.05 & -29.1 & 5.58 $\pm$ 2.77& 5.7\\
   J092634-262355 & 142.12 & 141.64 & -26.84 & -26.4 & 3.62 $\pm$ 1.89 & 2.76\\
   J092902-293017 & 142.15 & 142.26 & -29.56 & -29.5 & 5.8 $\pm$ 3.01& 7.53\\
    J093800-291244  & 144.12 & 144.5 & -28.3 & -29.21 & 11.67 $\pm$ 4.43& 12.01\\
   J093959-330710* & 144.88 & 145.0 & -33.03 & -33.12 & 4.76 $\pm$ 2.34& 3.42\\
    J094401-305244* & 146.06 & 146.0 & -30.92 & -30.88 & 3.3 $\pm$ 1.51 & 3.26\\
    J094649-272049*& 146.53 & 146.71 & -27.74 & -27.35 & 2.04 $\pm$ 1.43& 1.95\\
   J095003-330824 & 147.47 & 147.51 & -33.19 & -33.14 & 4.08 $\pm$ 1.90& 4.2\\
    J094953-251138  & 147.48 & 147.47 & -25.19 & -25.19 & 5.31 $\pm$ 4.21& 11.47\\
     J095433-305326* &148.68 & 148.64 & -30.79 & -30.89 & 6.51 $\pm$ 3.21& 6.65\\
   J095804-290408* & 149.47 & 149.52 & -29.0 & -29.07 & 8.44 $\pm$ 5.2& 8.26\\
  J100439-321639 &151.14 & 151.16 & -32.25 & -32.28 & 4.72 $\pm$ 2.3& 5.03\\
  J100910-285552 & 152.2 & 152.29 & -28.9 & -28.93 & 5.13 $\pm$ 2.35& 5.7\\
  J100855-301114* & 152.34 & 152.23 & -30.2 & -30.19 & 5.09 $\pm$ 3.11& 5.01\\
   		J101329-283118 & 153.36 & 153.37 & -28.51 & -28.52 & 5.76 $\pm$ 2.34& 5.73\\  
  J101348-315323 & 153.45 & 153.45 & -31.88 & -31.89 & 5.84 $\pm$ 2.87& 7.14\\
  J102056-321100* & 154.52 & 155.24 & -31.72 & -32.18 & 5.97 $\pm$ 2.56& 5.21\\
  J102011-324533 & 155.1 & 155.05 & -32.68 & -32.76 & 7.13 $\pm$ 3.56 & 8.95\\
  J102020-333556* & 155.22 & 155.09 & -33.6 & -33.6 & 6.12 $\pm$ 3.29 & 6.0\\	
  J102633-294054* & 156.94 & 156.64 & -29.65 & -29.68 & 1.44 $\pm$ 1.04 & 1.62\\
  J103312-341842 & 158.27 & 158.3 & -34.31 & -34.31 & 7.26 $\pm$ 3.67& 12.12\\
  J103704-334313* & 159.32 & 159.27 & -33.6 & -33.72 & 7.25 $\pm$ 4.21 & 5.8\\
  J103723-290749 & 159.36 & 159.35 & -29.25 & -29.13 & 4.37 $\pm$ 2.13 & 3.57
\end{tabular}
\end{table*}

\end{document}